\documentclass[12pt,prd,draft,nofootinbib]{revtex4}
\usepackage{amsfonts}
\usepackage{amssymb}
\usepackage{amsmath}
\usepackage{bm}
\usepackage{latexsym}
\newcommand{\beq}{\begin{eqnarray}}
\newcommand{\eeq}{\end{eqnarray}}
\newcommand{\be}{\begin{equation}}
\newcommand{\ee}{\end{equation}}
\def\la{\mathrel{\mathpalette\fun <}}
\def\ga{\mathrel{\mathpalette\fun >}}
\def\fun#1#2{\lower3.6pt\vbox{\baselineskip0pt\lineskip.9pt
\ialign{$\mathsurround=0pt#1\hfil ##\hfil$\crcr#2\crcr\sim\crcr}}}
\newcommand{\veX}{\mbox{\boldmath${\rm X}$}}
\newcommand{{\SD}}{\rm SD}
\newcommand{\pp}{\prime\prime}
\newcommand{\vePc}{\bm{P}}
\newcommand{\veY}{\mbox{\boldmath${Y}$}}
\newcommand{\vex}{\bm{x}}
\newcommand{\vey}{\bm{y}}
\newcommand{\vez}{\bm{z}}
\newcommand{\ver}{\bm{r}}
\newcommand{\vesig}{\bm{\sigma}}
\newcommand{\vedelta}{\mbox{\boldmath${\delta}$}}
\newcommand{\veP}{\bm{P}}
\newcommand{\vep}{\bm{p}}
\newcommand{\veq}{\mbox{\boldmath${q}$}}
\newcommand{\veS}{\mbox{\boldmath${S}$}}
\newcommand{\veL}{\bm{L}}
\newcommand{\vel}{\bm{\boldmath${l}$}}
\newcommand{\veR}{\mbox{\boldmath${R}$}}
\newcommand{\ves}{\mbox{\boldmath${s}$}}
\newcommand{\vek}{\mbox{\boldmath${k}$}}
\newcommand{\ven}{\mbox{\boldmath${n}$}}
\newcommand{\veu}{\mbox{\boldmath${u}$}}
\newcommand{\vev}{\mbox{\boldmath${v}$}}
\newcommand{\veh}{\mbox{\boldmath${h}$}}
\newcommand{\verho}{\mbox{\boldmath${\rho}$}}
\newcommand{\vexi}{\mbox{\boldmath${\xi}$}}
\newcommand{\veta}{\mbox{\boldmath${\eta}$}}
\newcommand{\veB}{\mbox{\boldmath${B}$}}
\newcommand{\veH}{\mbox{\boldmath${H}$}}
\newcommand{\veE}{\mbox{\boldmath${E}$}}
\newcommand{\veJ}{\mbox{\boldmath${J}$}}
\newcommand{\veal}{\mbox{\boldmath${\alpha}$}}
\newcommand{\vegam}{\mbox{\boldmath${\gamma}$}}
\newcommand{\vepar}{\mbox{\boldmath${\partial}$}}
\newcommand{\llan}{\langle\langle}
\newcommand{\rran}{\rangle\rangle}
\newcommand{\lan}{\langle}
\newcommand{\ran}{\rangle}
\begin{document}
\title{ Decay constants of the  heavy-light  mesons from  the field
correlator method }
\author{A.M. Badalian$^a$, B.L.G. Bakker$^b$, and Yu.A. Simonov$^a$\\
$^a$ State Research Center\\
Institute of Theoretical and Experimental Physics,\\
Moscow, 117218 Russia\\
$^b$ Department of Physics and Astronomy,\\
Vrije Universiteit, Amsterdam}
\date{\today}

\begin{abstract}
Meson Green's functions and decay constants $f_{\Gamma}$ in
different channels $\Gamma$ are calculated using the Field
Correlator Method. Both, spectrum and $f_\Gamma$, appear to be
expressed only through universal constants: the string tension
$\sigma$, $\alpha_s$, and the pole quark masses. For the $S$-wave
states the calculated masses agree with the experimental numbers
within $\pm 5$ MeV.  For the $D$ and $D_s$ mesons the values of
$f_{\rm P} (1S)$ are equal to 210(10) and 260(10) MeV,
respectively, and their ratio $f_{D_s}/f_D$=1.24(3) agrees with
recent CLEO experiment. The values $f_{\rm P}(1S)=182,\,216,\,438$
MeV are obtained for the $B$, $B_s$, and $B_c$ mesons with the
ratio $f_{B_s}/f_B$=1.19(2) and $f_D/f_B$=1.14(2). The decay
constants $f_{\rm P}(2S)$ for the first radial excitations as well
as the decay constants $f_{\rm V}(1S)$ in the vector channel are
also calculated. The difference of about 20\% between $f_{D_s}$
and $f_D$, $f_{B_s}$ and $f_B$ directly follows from our
analytical formulas.
\end{abstract}

\maketitle

\section{Introduction}
\label{sect.1}

The decay constants $f_{\rm P}$ in the pseudoscalar (P) channel, being
important characteristics of mesons, in many cases can be directly
measured in experiment, and therefore they can provide a precise manner to
compare different theoretical approaches and check their accuracy.
During the last decade the constants $f_{\rm P}$ have been studied by
many authors in potential models \cite{1,2,3,4,5,6,7,8}, in the QCD sum rule
method \cite{9,10}, in lattice simulations \cite{11,12,13,14},
as well as in experiment \cite{15,16,17,18,19}.(The papers
\cite{2,3,4,5,6,7} contain references and  a summary of numerous
calculations of decay constants).

The present article is devoted to the systematic derivation of the
meson Green's functions in QCD and study of the decay constants for
channels with arbitrary quantum numbers $\Gamma$, of which we
specifically consider P and vector (V) channels. For the decay constant
$f_\Gamma$ transparent analytical expressions will be obtained and in
particular, using those, the difference between $f_{D_s}$ and $f_D$,
$f_{B_s}$ and $f_B$ can be easily explained.

This paper is an improvement and extension of the earlier paper
\cite{20} devoted to the heavy-light (HL) pseudoscalars.
Ref.~\cite{20} appeared before the systematic formulation of The Field
Correlator Method (FCM) \cite{21}, in particular, before the derivation
of the string Hamiltonian \cite{22}, therefore some steps in \cite{20}
were not accurately proved.  In this paper we give a consistent and
general treatment of the meson Green's function and its spectral
properties.  The main problem, which one encounters when addressing the
spectral properties in QCD, is the necessity to include quantitative
nonperturbative (NP) methods, which  are responsible for  the main
dynamical phenomena: confinement and chiral symmetry breaking (CSB).

In the  FCM, introduced in \cite{23}, one derives  the effective
Hamiltonian, which comprises both confinement and relativistic effects,
and contains only universal quantities: the string tension $\sigma$,
the strong coupling $\alpha_{\rm s}$, and  the current (pole) quark
masses $m_i$. We use here the pole quark masses which correspond to the
conventional current (Lagrangian) masses $\bar m_q(\bar m_q)$
\cite{16}. The simple local form of this Hamiltonian, which will be
called the string Hamiltonian (SH), occurs for objects with temporal
scales larger than the  vacuum gluon correlation length, $T_g \approx
0.2$ fm, i.e., it is applicable to all QCD bound states with  an
exception of toponium.  Explicit calculations of masses and wave
functions with the use of the SH have been done recently for light mesons
\cite{24}, heavy quarkonia \cite{25}, and heavy-light mesons
\cite{26}, and demonstrate  good agreement with experimental masses,
leptonic widths, and fine structure effects.

As compared to QCD sum rules and lattice QCD this method has an
essential advantage, because the radial and orbital excitations
can be considered in this approach on the same grounds as the ground
states.

The calculation of spectral coefficients, like decay constants
$f_\Gamma$, needs an additional step,namely, besides using the SH
the computation of all coefficients in the Green's function,
including the Dirac spinor structure etc.  Moreover, for $\pi$ and
$K$ mesons CSB is vitally important.  Recently the FCM was
extended to include the effects of CSB \cite{27},  where it was
shown that the phenomenon of CSB occurs due to confinement and two
characteristic parameters of CSB -- the chiral condensate and
$f_\pi$ -- were computed in terms of $\sigma$. The calculation of
$f_\pi$ and $f_K$ can be done using a simple extension of  general
expression (23), derived in Ref.~\cite{27}, while here we
concentrate on the calculations of the masses and $f_\Gamma$ of HL
mesons; our method also enables one to calculate $f_\Gamma $ for
excited states and here decay constants will be calculated for the
$2S$ states.

One  important technical problem, which is solved in this paper and
allows one to calculate $f_\Gamma$ in all channels, is the accurate
einbein reformulation of the Fock-Feynman-Schwinger Representation
(FFSR), or the world-line representation, where the dynamical quark
mass $\omega_q$ appears as an integration variable instead of the
proper time. The previous step in this direction \cite{22} has enabled
one to write only the string Hamiltonian, but the whole Green's
function was not attainable.  Below in the FFSR we derive the explicit
einbein form of the meson Green's function.

The paper is organized as follows: in Section~\ref{sect.2} the general
einbein form of the meson Green's function is presented, while in
Appendices \ref{app.A} - \ref{app.C} the details of the derivation are
given. In Section ~\ref{sect.3} the masses of the heavy-light pseudoscalar
mesons are considered. The decay constants are calculated in Section 4,
while  auxiliary variables are given in Appendices~\ref{app.D} and
\ref{app.E}.  Section~\ref{sect.5} is devoted to the approximations
used in our calculations  and Section~\ref{sect.6} contains our concluding
remarks.

\section{ The meson Green's function in the FFSR}
\label{sect.2}

Here  we derive the einbein form of the meson Green's function
written as the path integral in the FFSR~\cite{28,29,30}. We start
with the FFSR for the quark Green's function in the gluonic field
$A_\mu$, which contains both perturbative and NP contributions and
Euclidean space-time is assumed everywhere:
\begin{eqnarray}
 S(x,y) & = & (m+\hat D)^{-1}=(m-\hat D) (m^2-\hat D^2)^{-1}=
\nonumber \\
 & & = (m-\hat D) \int^\infty_0 ds(Dz)_{xy} e^{-K}
 \Phi_\sigma(x,y),
\label{1}
\end{eqnarray}
with
\begin{eqnarray}
 \hat D & = & D_\mu\gamma_\mu, \quad D_\mu =
 \frac{\partial}{\partial x_\mu}-ig A_\mu,
\nonumber \\
 K & = & m^2s+\frac{1}{4} \int^s_0 \left( \frac{dz_\mu}{d\tau}\right)^2
 d\tau,
\nonumber \\
 \Phi_\sigma (x,y) & = & P\exp \left(ig \int^x_y A_\mu dz_\mu\right)
 \exp \left(g\int^s_0 \sigma_{\mu\nu} F_{\mu\nu} d\tau\right),
\label{2aa}
\end{eqnarray}
where
\begin{eqnarray}
 (Dz)_{xy} & =& \lim_{N\to\infty}\prod^N_{k=1}\left. \frac{d^4\Delta
 z(k)}{(4\pi \varepsilon)^2}\right|_{ \sum_k\Delta z(k) =x-y},
 N\varepsilon =s,
\nonumber \\
 \left|_{a=b} \right. & \equiv& \int\frac{d^4p}{(2\pi)^4} e^{ip (a-b)}.
\label{eq.2ab}
\end{eqnarray}

In Eq.~(\ref{1}) the role of the evolution parameter is played by
the proper time $s$, whereas in $(Dz)_{xy}$ there is an
integration over the fourth component $z_4(\tau)$, ($0\leq \tau
\leq s),$ which is the Euclidean time of the particle. The crucial
point now is  to go over in Eq.~(\ref{1}) to the Euclidean time
$z_4\equiv t$ as an evolution parameter. To get rid of the proper
time $s$, one can use the so-called einbein method \cite{31},
which was applied to the FFSR in Ref.~\cite{22,30}, and here it is
developed and used for the correlator of the currents.

To this end  the so-called dynamical mass (variable) $\omega(t)$
can be introduced  via the relation  between the proper time
$\tau$ and  Euclidean time $t$:
\be
 d\tau = \frac{dt}{2\omega(t)},
 \quad \bar \omega =\frac{1}{s} \int^s_0 \omega (\tau) d\tau=
 \frac{1}{N} \sum^N_{k=1} \omega(k).
\label{3}
\ee
The  integrals in
Eq.~(\ref{1}) can now be identically rewritten as
\be
 ds (Dz)_{xy} \equiv (D^3z)_{ \vex\vey} (D\omega)
 = \lim_{N\to\infty} \prod^N_{k=1}\left.
 \frac{d^3\Delta z(k)}{ {2\bar\omega}l^3(k)}
 \right|_{\sum_k\Delta \vez =\vex-\vey}
 \frac{d\omega(k)}{l_\omega(k)},
\label{4}
\ee
where
\be
 l(k) =\sqrt{\frac{2\pi\Delta t}{\omega(k)}}, \quad
 l_\omega (k) =\sqrt{\frac{2\pi\omega(k)}{\Delta t}}, \quad
 N\Delta t = x_4 -y_4 \equiv T.
\label{5}
\ee
In Appendix~\ref{app.A} the representations (\ref{1}), (\ref{4}) are
illustrated by calculating the free quark propagator, where the meaning
of $\omega(k)$ and $\bar \omega$ in the momentum representation appears
to be very simple: $\bar \omega =\omega(k) = \sqrt{\vep^2+m^2}$.

We now turn to the meson (quark-antiquark) case and consider the
correlator $G_\Gamma (x)$ of the currents $j_\Gamma(x)$:
\begin{eqnarray}
 j_\Gamma(x) & = & \bar \psi_1 (x) \Gamma\psi_2 (x),
\nonumber \\
 \Gamma & = & t^a\otimes (1, \gamma_5, \gamma_\mu, i \gamma_\mu \gamma_5)
{\rm ~for ~ S,P,V,and ~A~ channels},
\nonumber \\
 t^a & = & \frac{\lambda^a}{2}, \quad tr (t^a t^b)=\frac12 \delta_{ab},
 \label{6}
\end{eqnarray}
and
\begin{eqnarray}
 G_\Gamma (x) & \equiv & \lan j_\Gamma (x) j_\Gamma (0)\ran_v
\nonumber \\
 & = & 4 N_c\int Y_\Gamma (D^3z)_{\vex 0} (D^3\bar z)_{\vex
0}(D\omega_1) (D\omega_2) \exp (-K_1-K_2) W_\sigma.
 \label{7}
\end{eqnarray}
In (\ref{7}) we have defined the new quantity $Y_\Gamma$:
\be
 4Y_\Gamma = tr_L(m_1-\hat D_1) \Gamma(m_2-\hat D_2)\Gamma \to tr_L
 (m_1+\omega_1\hat{\dot{z}}) \Gamma(m_2- \omega_2 \hat{\dot{\bar{z}}})\Gamma,
\label{8}
\ee
which can also be
written in the operator form:
\begin{equation}
 4Y_\Gamma= tr_L((m_1-i\hat p_1) \Gamma(m_2+i \hat p_2)\Gamma),
\label{eq.7a}
\end{equation}
with $p^{(i)}_\mu \;(\hat p_i=p_\mu\gamma_\mu)$ -- the  momentum
of particle $i$ as it is derived in Appendix~\ref{app.B} for the
convenience of the reader ($tr_L$ means the trace over Dirac
indices). The resulting expressions for $Y_\Gamma$ in the V, A, S,
and P channels are given below in Eqs.~(\ref{34}).

In (\ref{7}) the symbol $W_\sigma = \Phi_{\sigma_1} (x,y)
\Phi_{\sigma_2}(y,x)$ stands for the average value of the Wilson
loop with the insertions of the operator
\begin{equation}
 \Lambda^{(1)}_\sigma\equiv \exp \left(g\sigma^{(1)}_{\mu\nu} \int^s_0
F_{\mu\nu} d\tau\right) =\exp \left( g \sigma^{(1)}_{\mu\nu} \int^T_0
F_{\mu\nu} (z(t_1))\frac{dt_1}{2\omega_1(t_1)}\right)
\label{eq.8a}
\end{equation}
for the quark line and of the operator $\Lambda^{(2)}_\sigma$ for the
antiquark line:
\begin{equation}
 \Lambda^{(2)}_\sigma \equiv \exp \left(-g\sigma^{(2)}_{\mu\nu}
 \int^s_0 F_{\mu\nu} d\tau\right) = \exp \left(-g \sigma^{(2)}_{\mu\nu}
 \int^T_0 F_{\mu\nu}{(\bar z (t_2))}
\frac{dt_2}{2\omega_2(t_2)}\right).
\label{eq.8b}
\end{equation}
Since in HL mesons we  will consider spin-effects as a perturbation, in
the first approximation  both factors $\Lambda^{(1)}_\sigma $ and
$\Lambda_\sigma^{(2)}$ are replaced by 1 and $W_\sigma$  is
simplified.

In Appendix~\ref{app.C} one can find all explicit steps for the derivation
of the correlator (\ref{7}) in the  simplest case when the gluon
interaction is absent, i.e., for $W_\sigma \equiv 1$. One can see there
that the quark-loop contribution is reconstructed with the correct
coefficients. Now we turn to the case when the NP interaction is
included in $W_\sigma$.

In general, $W_\sigma$ contains all effects of the interaction
which include: (i) the perturbative static gluon exchange; (ii)
the radiative corrections to $G_\Gamma$ in the form of the
operator anomalous dimension and corrections to $f_\Gamma$; (iii)
the NP contributions to $G_\Gamma$. To calculate all of them one
can use the background perturbation theory \cite{32,33} and the
FCM. For all hadrons of interest to us (with a size larger than
$T_g \approx 0.2$ fm)  the use of the FCM reduces to the
appearance of the area-law factor in $W_\sigma$, which is
accompanied by Coulomb, radiative, and spin-dependent factors.
(For more discussion see the reviews \cite{30,34}). As was shown
in Refs.~\cite{33,34} this produces (apart from radiative
corrections) the local interaction $\hat V (r) $, which for not so
large angular momentum ($L\leq 4 $) can be presented  as $W_\sigma
= \exp (-\int_0^T dt \hat V)$, where the interaction,
\be
 \hat V(r) = V_0(r) + V_{\rm SD} + \Delta V_{\rm string} +V_{\rm SE},
\label{9}
\ee
contains the static potential:
\be
 V_0 (r) = \sigma r -\frac43 \frac{\alpha_{\rm st}(r)}{r},
\label{9a}
\ee
and the spin-dependent part $V_{\rm SD} (r)$ given by:
\begin{equation}
 V_{\rm SD} = V_{\rm SS}+ V_{\rm LS} +V_{\rm T},
\label{eq.10a}
\end{equation}
with spin-spin, spin-orbit, and tensor terms; the self-energy
contribution $V_{\rm SE}$ \cite{35}, and also  a  ``string correction"
(occurring only for the states with $L\neq 0$ \cite{22,24}). For the
$S$-wave mesons, considered here,
\be
 \hat  V(r)=V_0(r) +V_{\rm SS}+ V_{\rm SE},
\label{11}
\ee
while for the spin-averaged masses only two terms are left in  the
potential, $V(r) =V_0(r)+ V_{\rm SE}$. In Section~\ref{sect.5} we shall
also take into account radiative corrections and the operator anomalous
dimension.

One can now rewrite $G_\Gamma$ in Eq.~(\ref{7}), separating c.m. and
relative distance coordinates,
\begin{equation}
 \veta= \Delta \vez_1 -\Delta
 \vez_2, \quad \verho=\frac{\omega_1\Delta \vez_1 +\omega_2 \Delta
\vez_2}{\omega_1+\omega_2}, \quad
 \omega_{\rm r} = \frac{\omega_1 \omega_2}{\omega_1+\omega_2},
\label{10}
\ee
where all coordinates are labelled with the index $k$ as in Eq.~(\ref{5}).

Integrating out the c.m. coordinate $d\verho$ and
$d\omega_+=\omega_1+\omega_2$, as shown in Appendix~\ref{app.C},
one arrives at the path integral in the relative coordinate
$d\veta$, which can be expressed through the SH. Indeed, from the
path integral formalism \cite{30,36} it is known that a general
equivalence relation holds:
\be
 \int \frac{(D^3\eta)_{xy}}{(2\pi\Delta t/ \omega_{\rm r} (k))^{3/2}}
 e^{-\sum_k \left( \frac{\omega_{\rm r} (k) \eta^2(k)}{2\Delta t} +
\hat V(k) \Delta t\right)}= \langle x|e^{-\hat HT} |y\ran,
\label{17}
\ee
where
\be
 \hat H = \frac{\vep^2_\eta}{2\omega}_{\rm r} + \hat V(\eta), \quad
 \vep_\eta =\frac{1}{i} \frac{\partial}{\partial\veta}.
\label{18}
\ee
Taking into account the integral (see Appendix~\ref{app.C}):
\be
 \int \frac{2d \omega_{\rm r}(k)}
 {\sqrt{\omega_{\rm r} (k)}\sqrt{\frac{2\pi}{\Delta t}}}
 e^{-2 \omega_{\rm r} (k) \Delta t-
 \frac{(\bm{p}^2+m^2)\Delta t}{2 \omega_{\rm r} (k)} } =
 e^{-2\sqrt{\vep^2+m^2}\Delta t},
\label{19}
\ee
one obtains the important relation:
\be
 \int G_\Gamma(x) d^3\vex = N_c \left\lan 0 \left|\frac{ Y_\Gamma}{\bar
\omega_1\bar\omega_2}  e^{-\hat  H T}\right| 0\right\ran,
\label{20}
\ee
where the Hamiltonian is
\be
 \hat H = \sqrt{\vep^2+m^2_1} + \sqrt{\vep^2+m^2_2}+\hat V (r).
\label{21}
\ee
Then  with the use of the spectral expansion the expression
(\ref{20}) can be presented as
\be
 \left\lan 0 \left|\frac{ Y_\Gamma}{\bar \omega_1\bar\omega_2}
 e^{-\hat  H T}\right| 0\right\ran
 = \sum_n\frac{\lan Y_\Gamma\ran_n |\varphi_n(0)|^2}
 {\lan\bar \omega_1\ran_n \lan \bar \omega_2\ran_n}  e^{-M_nT},
\label{18a}
\ee
where $\varphi_n$ and $M_n$ are the eigenfunction (e.f.) and the
eigenvalue (e.v.) of the  Hamiltonian (\ref{21}).  It follows from the
extremum conditions (\ref{D.7}) that the variables $\bar \omega_1$ and
$\bar \omega_2$ in  Eq.~(\ref{20}) are defined as the operators,
\be
 \bar \omega_i= \sqrt{\vep^2+m^2_i},
\label{22}
\ee
while in Eq.~(\ref{18}) $\lan \bar\omega_i\ran_n$ is the matrix element over
this operator for the $nL$ state.

On the other hand for the l.h.s. of Eq.~ (\ref{20}) one can also use
the conventional spectral decomposition:
\begin{eqnarray}
 \int G_\Gamma(x) d^3 \vex & = &
 \sum_n \int d^3 \vex \lan 0 | j_\Gamma| n\ran \lan n |j_\Gamma|0\ran
 e^{i\vePc \cdot \vex -M_nT}\frac{d^3\vePc }{2M_n(2\pi)^3}
\nonumber \\
 & = & \sum_n \varepsilon_\Gamma \otimes \varepsilon_\Gamma
 \frac{(M_n f_\Gamma^n)^2}{2M_{n}} e^{-M_nT}.
\label{23}
\end{eqnarray}
Here we have used the  standard definition for  $f^n_\Gamma \equiv
f_\Gamma (nS)$:
\be
 \lan 0|j_\Gamma| n, \vePc=0\ran = \varepsilon_\Gamma
 M_n f^n_\Gamma,
\label{24}
\ee
where $\varepsilon_\Gamma =\varepsilon_\mu^{(k)}$ for V and A channels,
$\varepsilon_\Gamma =1$ for S and P channels, while the polarization
vector $\varepsilon_\mu^{(k)}$ satisfies the normalization condition:
\be
 \sum_{k=1,2,3} \varepsilon_\mu^{(k)} (q) \varepsilon_\nu^{(k)} (q) =
 \delta_{\mu\nu}-\frac{q_\mu q_\nu}{q^2}.
\label{25}
\ee
Inserting the expression (\ref{18a}) into (\ref{20}) and using the relation
(\ref{23}), one obtains the decay constant $f^n_\Gamma$ in the
channel $\Gamma$ for  the $nS$ state:
\be
(f_\Gamma^n)^2=\frac{2N_c \lan Y_\Gamma\ran | \varphi_n
(0)|^2}{\lan \bar \omega_1\ran_n \lan \bar\omega_2\ran_n
M_n}=\frac{6|\varphi_n(0)|^2}{M_n} \frac{\lan Y_\Gamma\ran }{\lan
\bar \omega_1\ran_n \lan \bar \omega_2\ran_n }.
\label{26b}
\ee

To derive the relations (\ref{18a}) and (\ref{26b}) we have used the
essential property of the SH Eq.~(\ref{21}) that the average value
$\lan \omega_i (k)\ran $ of the operator $\omega_i(k)$ is equal to the
average of the operator $\sqrt{ \vep^2 +m^2_i}$, and hence the average
$\bar \omega_i =\frac{1}{N} \sum^N_{k=1} \omega_i (k)$, denoted as
$\lan \bar \omega_i \ran _n$ is
\begin{equation}
 \lan \bar \omega_i \ran_n = \lan \omega_i (k) \ran_n = \lan
\sqrt{\vep^2 + m^2 }\ran_n,
\label{eq.23a}
\end{equation}
where the average is assumed to be taken over the eigenstate $\varphi_n (r)$:
\begin{equation}
 \lan \bar \omega_i \ran_n =
 \lan \varphi_n | \sqrt{\vep^2 +m^2_i} |\varphi_n\ran.
\label{eq.23b}
\end{equation}

In the most general case this average may differ from the
path-integral average $\lan\bar \omega_i\ran$ in Eqs.~(\ref{3}),
(\ref{4}), (\ref{7}).  However, the analysis of this problem, done
in Ref.~\cite{37}, shows that the difference between the two
definitions is small ($\la 3\%$ for the lowest states) and we
assume here that the same accuracy holds for our basic relation
(\ref{26b}).

In Eq.~(\ref{26b}) the mass of the $^1 S_0$-wave meson $M_n(^1 S_0)=
M_{\rm on} -\frac34 \Delta_{\rm HF}  + \Delta_{\rm SE},$ where $M_{\rm
on}$ is the e.v. of the Hamiltonian (\ref{21}) with  the static
interaction (\ref{9a}):
\begin{eqnarray}
 H_0 & = & \sqrt{m^2_1+\vep^2} + \sqrt{m^2_2+\vep^2} + V_0 (r),
\nonumber \\
 H_0\varphi_n & = & M_{\rm on}\varphi_n .
\label{27}
\end{eqnarray}
The general  expression of the self-energy correction to the meson mass
$M(nS)$ is calculated in \cite{35} (see Appendix~\ref{app.E}) (it comes
from the NP contributions to the quark and antiquark  mass):
\be
 \Delta_{\rm SE} =\sum_{i=1,2} \left\{\left( - \frac{1.5 \sigma \eta_{f_i}}
 {\pi \lan\bar \omega_i\ran}\right) +
 \frac{\sigma^2}{4\lan\bar\omega_i\ran [m_i+ \lan\bar \omega_i\ran/2
 +\varepsilon(\tilde \omega)]^2}\right\} ,
\label{25a}
\ee
where the factor $\eta_f(i)$ depends on the flavor of the $i$-th
quark (antiquark) and its  analytical expressions (\ref{E.2})  are
deduced in Ref.~\cite{35}. For the  $u(d)$, $s$, $c$, and $b$
quarks the values
\be
 \eta_{u (d)} =1.0, \quad \eta_s =0.65,\quad \eta_c =0.35,\quad
 \eta_b=0.03
\label{26a}
\ee
are obtained in Appendix~\ref{app.E}\footnote{Note, that these values
of $\eta_{f(i)}$ in (\ref{26a}) are not fitting parameters, but calculated
through the same input $m_i$ and $\sigma$}.

It is worthwhile to notice that for the  $b$ quark $\eta_b$ is small
and  its  contribution to $\Delta_{\rm SE}$ is small ($\sim 1$ MeV) and
can be neglected. Therefore, for the $B$, $B_s$, and $B_c$ mesons
 we have to use in Eq.~(\ref{25a}) only a contribution which comes from
 the lighter quark (antiquark) $q_1 (\bar q_1)$ denoted later by  the
index 1. For the $D$ and $D_s$ mesons both terms (with $i=1$ and
$i=2$) are taken into account, although the contribution from the
c-quark term is small, around -20 MeV.

Thus, the scheme of our calculations of $f_\Gamma$ is as follows:

1. First, one calculates the e.v. $M_{\rm on}$ of the SH (\ref{21})
with the static interaction (\ref{9a}) and then takes into account the
self-energy (\ref{25a}) and the HF corrections:
\begin{equation}
 M_n(^3 S_1) = M_{\rm on} +\Delta_{\rm SE} +\frac14 \Delta_{\rm HF},
 \quad M_n (^1 S_0) = M_{\rm on} + \Delta_{\rm SE} -\frac34 \Delta_{\rm HF}.
\label{eq26c}
\end{equation}

2. The values $\lan\bar \omega_1\ran$ and $\lan\bar \omega_2\ran$ in
Eqs~(\ref{26b}) and (\ref{25a}), are  the matrix elements (m.e.) of the
kinetic energy term defined in Eq.~(\ref{22}).

3. The factor $\lan Y_\Gamma\ran$ (in the channel $\Gamma$) can be
computed in terms of the momenta of a quark and an antiquark, or in the
c.m. system in terms of the relative momentum $\vep$, with the  following
results for the m.e. $\lan Y_\Gamma\ran$ (see  Appendix~\ref{app.B}):
\begin{eqnarray}
 \lan Y_{\rm V}\ran & = & m_1m_2+\lan \bar \omega_1\ran\lan\bar\omega_2\ran
+\frac13\lan \vep^2\ran,
\nonumber \\
 \lan Y_{\rm P}\ran & = & m_1m_2+\lan\bar \omega_1\ran\lan\bar\omega_2\ran
 -\lan \vep^2\ran =\lan Y_{A_4}\ran
. \label{34}
\end{eqnarray}
Here we use for the operators $Y_\Gamma$ the notations:
\begin{eqnarray}
 \hat Y_{\rm V} & = & \frac{1}{3}\sum_i
 tr [(m_1-\hat D_1) \gamma_i (m_2-\hat D_2)\gamma_i],
\nonumber \\
 \hat Y_{\rm P} =\hat Y_{A_4} & = &-[tr (m_1-\hat D_1) \gamma_4\gamma_5
(m_2-\hat D_2)\gamma_4\gamma_5].
\label{37}
\end{eqnarray}
In the case of the P channel with  both $m_1,m_2\to 0$ due to CSB
there appears an additional mass term in Eq.~(\ref{34}), which can
be computed through field correlators.  (The $\pi$ and $K$ mesons
will be considered later \cite{27}).

For the calculations of $\lan Y_\Gamma\ran $ one needs also to know the
m.e. $\lan  \vep^2\ran_{nS}$ and the w.f. at the origin, $\varphi_n(0)=
R_n(0)/\sqrt{4\pi}$; they are given in Appendix~\ref{app.D}.

4. Our calculations are done with the relativistic SH (\ref{21})
which was derived in  einbein approximation (EA), therefore the
w.f. at the origin must also be calculated with the use of the EA
(see Appendix~\ref{app.D}), which provides an accuracy of $\la$
5\% \cite{26,37}. In the nonrelativistic limit, $m_i\gg
\sqrt{\sigma}$, one can easily find  that $\lan \bar
\omega_i\ran\approx m_i$,  while $\lan \vep^2\ran \sim O(\sigma)$
can be neglected in Eqs.~(\ref{34}), and therefore
\begin{eqnarray}
 \lan  Y_{\rm V}\ran_{\rm NR} & \approx & 2m_1m_2 + O(\sigma),\quad
 \lan Y_S\ran_{\rm NR}\approx O(\sigma),
\label{38}
\nonumber \\
 \lan Y_{A_i}\ran_{\rm NR} & \approx & O(\sigma),
\quad \lan Y_{A_4}\ran_{\rm NR}= 2m_1m_2 +O(\sigma),
\label{39}
\nonumber \\
 \lan Y_{\rm P}\ran_{\rm NR}  & =  & 2m_1m_2 +O(\sigma).
\label{40}
\end{eqnarray}
Thus in  the nonrelativistic limit for $f_\Gamma^{(n)}$ in the V and P
channels one obtains the well-known result \cite{1}:
\be
 (f^n_\Gamma)^2_{\rm NR} =\frac{4N_c}{M_n} |\varphi_n (0)|^2 \quad
 (\Gamma=V,P),
\label{eq.36}
\ee
while in the $S$ channel $f_S\to 0$.

5. As a final step one needs to compute the radiative corrections
to $f^n_\Gamma$, which  come from the short-distance (large
momentum) perturbative gluon contributions. Neglecting
interference terms they can be written as  in \cite{20,30},
\be
 \lan W_\sigma\ran = \lan W_{\rm OGE} \ran \lan W_{\rm nonpert}\ran
\label{45}
\ee
with
\be
 \lan W_{\rm OGE}\ran = Z_m \exp \left( -\frac{4}{3\pi} \int \int
 \frac{dz_4 dz'_4 \alpha_{\rm s} (z-z')}{(z-z')^2} \right),
\label{46}
\ee
where $Z_m$ is a regularization factor. After separating  the Coulomb
interaction in $\hat H$ in this  way, one gets the correction to $\lan
W_\sigma\ran$, and  $f^2_\Gamma$   can be written  in the form:
\be
 f^2_\Gamma\to \xi_\Gamma  f^2_\Gamma, \quad
 \xi_\Gamma  =1+c_\Gamma\alpha_{\rm s} + O(\alpha^2_{\rm s}).
\label{47}
\ee
Another important contribution from perturbative gluon exchanges (GE)
is the account of Asymptotic Freedom  (AF) in the coupling
constant $\alpha_{\rm s}$ in (\ref{46}), which is especially important
for  the value  of $\varphi_n(0)$ in the $S$-wave channels. In our
calculations we use the GE interaction where in the strong coupling
$\alpha_B(r)$ the AF behavior is taken into account. Nevertheless, it is
of interest to compare the w.f. at the origin
with and without AF behavior in the GE term, introducing the factor
$\rho_{AF} =\left | \varphi_n^{({\rm AF})}(0)/ \varphi_n (0) \right|^2$,
which appears to be around 0.80 for the $1S$ states \cite{20}.  Then
$f^2_\Gamma$ (with AF taken into account) can be expressed through
$\tilde f^{2}_\Gamma$ (where AF is neglected) as $f^2_\Gamma = \tilde
f^2_\Gamma ~~\xi_\Gamma\rho_{\rm AF}.$

6. We conclude this section with the discussion of the input
parameters $m_i$, $\alpha_{\rm s}$ and $\sigma$.  We take
$\sigma=0.18$ GeV$^2$ for all HL mesons (as in light mesons and in
heavy quarkonia  \cite{38}); $m_i$ are the conventional pole
masses which are defined through the Lagrangian (current) masses
in the $\overline{MS}$-scheme (see \cite{16,39} and references
therein):
\be
 m_i =\bar m_{\overline{MS}} ( \bar m_{\overline{MS}}) \left\{
 1+ \frac43 \frac{\alpha_{\rm s} (\bar m_{\overline{MS}})}{\pi} +\eta_2
 \left(\frac{\alpha_{\rm s}}{\pi}\right)^2 + O({\alpha_{\rm s}})^3\right\}.
\label{49a}
\ee
We use here the pole masses presented in Table~\ref{tab.1}. They
correspond to the conventional current masses $\bar m_c=1.18$ GeV and
$\bar m_b=4.20$ GeV while for the strange quark  the pole mass, $m_s
=180$ MeV is taken at the scale $\mu\approx 0.5$ GeV $(r_0\approx 0.5$
fm) and therefore it should  be larger than the standard $m_s$(2 GeV)
$\approx 100$ MeV \cite{16}. For the $u$ and $d$ quarks $m_u=5$ MeV,
$m_d=8$ MeV are taken.

\begin{table}
\caption{The pole quark masses $m_q$ (pole) (in GeV), used in this
paper, and the constituent masses $m_q^{\rm C}$ from \cite{2,4,5}.\label{tab.1}}
\begin{center}
\begin{tabular}{|l|l|l|l|l|l|}\hline
~quark &~~ b &~~c &~~ s &~~ u &~~ d \\
\hline
~$m_q$ (pole)  this paper~&~4.78~ &~1.40~ &~0.180~&~0.005~&~0.008 ~\\
\hline
~$m_q$  from [4]&~4.655~&~1.511~&~0.216~&~0.071~&~0.071~\\
\hline
~$m_q^{\rm C}$ from [2]~&~4.977~&~1.628~&~0.419~&~0.220~&~0.220~\\
\hline
~$m^{\rm C}_q$ from [5]~&~5.158~&~1.755~&~0.535~&~0.371~&~0.377~\\
\hline
\end{tabular}
\end{center}
\end{table}

One can see the essential difference between the input masses in our SH
and the constituent masses used in the relativistic instantaneous
Bethe-Salpeter Method \cite{5}, and also in the spinless Salpeter
equation \cite{2}. The $b$, $c$, and $s$ quark masses from \cite{4},
where the relativistic Dirac Hamiltonian is used for a light quark, are
not large and  can be considered as the pole masses, while $m_u=m_d=71$
MeV seem to be too large.

7. The coupling $\alpha_{\rm st}(r)$ in the GE term is taken here
as the vector coupling $\alpha_B(r)$ in background perturbation
theory (in two-loop approximation) from \cite{39}. At large
distances this theory predicts saturation: $\alpha_{B} (r)\to
\alpha_{\rm crit}=$ const (this coupling $\alpha_B(r)$ is in good
agreement with lattice potential at small distances \cite{40}):
\begin{eqnarray}
 \alpha_B(r) & = & \frac{2}{\pi} \int^\infty_0 dq \frac{\sin qr}{q}
 \alpha_B (q),
\label{49b}
\nonumber \\
 \alpha_B(q) & =& \frac{4\pi}{\beta_0 t_B}
 \left\{ 1- \frac{\beta_1}{\beta_0^2} \frac{\ln t_B}{t_B} \right\}, \quad
 t_B =\ln \frac{q^2+M^2_B}{\Lambda^2_B}.
\label{49c}
\end{eqnarray}
Here the background mass $M_B =1$ GeV can be expressed through the
$\sqrt\sigma$ \cite{38}, while the QCD constant $\Lambda_B$ is
given by
\be
 \Lambda_B = \Lambda_{\overline{MS}} \exp \left( \frac{\frac{31}{3}
 -\frac{10}{9} n_f}{2\beta_0} \right).
\label{eq.49d}
\ee
It is of interest to notice that in our calculations of HL meson
properties the value $n_f=3$ is strongly preferable.

In this way the problem is uniquely defined and no fitting parameters
are introduced. The resulting masses and decay constants of HL mesons
appear to be unbiased  theoretical predictions which will be compared
with recent experiments, lattice data, and other theoretical
predictions.

\section{Masses of heavy-light mesons}
\label{sect.3}

In the relativistic SH a correct choice of the static interaction
$V_0(r)$, which defines the e.v. of the unperturbed Hamiltonian
$H_0$ (\ref{27}), is of great importance. Here, for all HL mesons
we take the static potential from \cite{38,39}:
\be
 V_B(r) =\sigma r - \frac43 \frac{\alpha_B(r)}{r}
\label{3.1}
\ee
with $\sigma=0.18 $ GeV$^2$, the number of flavors $n_f=3$,  the
vector coupling $\alpha_B(r)$ (\ref{49b}) also contains
\be
 \Lambda_B (n_f=3) =360~{\rm MeV},\quad M_B=1.0~{\rm GeV}.
\label{3.2}
\ee
For this choice the saturated (critical) value of the vector coupling,
reached at large $r$, is equal $\alpha_{\rm crit} (n_f =3) =0.495$,
while at $r\approx r_0 =0.5$ fm, $\alpha_B(r_0)\cong 0.43$.

It is very convenient to start the calculations with the spin-averaged
masses $M_{\rm cog} (nL)$,
\begin{eqnarray}
 M_{\rm cog} (nS) & = & M_0 (nS) +\Delta_{\rm SE} (nS),
\nonumber \\
 M_{\rm cog} (nL) & =& M_0 (nL) +\Delta_{\rm SE} (nL) + \Delta_{\rm str} (nL)
 \quad (L\neq 0),
\label{3.3}
\end{eqnarray}
since these masses do not depend on any additional parameter. In
contrast to $M_{\rm cog}$ the masses of the singlet and triplet states
depend also on the parameters defining the HF interaction, or for states
with $L\neq 0$ the masses $M(J^{PC})$ depend also on the coupling
$\alpha_{\rm FS}(\mu_{\rm FS})$, which defines the fine-structure
splittings, and these couplings are not well determined. In
Eq.~(\ref{3.3})$M_0(nL)$ is the e.v. of Eq.~(\ref{27}) (see
Tables~\ref{tab.2},\ref{tab.5}).The self-energy term $\Delta_{\rm SE}(nL)$ (\ref{25a}) and the string
correction $\Delta_{\rm str} (nL)= \lan H_{\rm str} \ran $ (\ref{D.12})
are defined by analytical expressions and discussed in Appendices~\ref{app.D}
and \ref{app.E}.

\begin{table}
\caption{The spin-averaged masses $M_{\rm cog} (1S)$ (in MeV)$^{a)}$.
\label{tab.2}}
\begin{center}
\begin{tabular}{|l|l|l|l|l| }\hline
Meson multiplet& $M_0 (1S)$& $\Delta_{\rm SE}$& $M_{\rm cog}(1S)$&
$M_{\rm cog}(\exp)^{b)}$\\
\hline
$D - D^*$&~2139&~$-164$ &~1975 &~1974.8(4)\\
$D_s - D^*_s$ &~2177 &~$-105$ &~2072 &~2076.0(6)\\

$B - B^*$ &~5433&~$-120$ &~5313 &~5313.5(6)\\
$B_s - B^*_s$ &~5468 &~$-72$ &~5396 &~5400.7(32)\\
$B_c - B^*_c$ &~6332 &~$-17$ &~6315 &~~~~~ $-$\\
\hline
\end{tabular}

$^{a)}$ The $\Delta_{\rm SE}$ are calculated in Appendix~\ref{app.E}.

$^{b)}$ The experimental numbers are taken from PDG \cite{16} and
\cite{41}.
\end{center}

\end{table}

The calculated  $M_{\rm cog} (1S)$, as seen from
Table~\ref{tab.2}, agree with the experimental numbers with an
accuracy better than 5 MeV. It is of interest to notice that the
difference $M_{\rm cog} (D^+_s) - M_{\rm cog}(D^+)$, which in
experiment is 99 MeV, is only partly due to dynamical reasons:
$M_0 (D_s) - M_0(D)=38$ MeV, but mostly occurs due to the
difference in the SE contributions, equal to 60 MeV.

With the use of our number  $M_{\rm cog} (B_c) =6315$ MeV  and
$M(B_{c})_{\rm exp} =6275(7)$ MeV \cite{41} one  can predict  the
mass of  the  vector state $B_c^* (1^3S_1)$:
\be
 M(B_c^*)= 6328(7)~{\rm MeV,~if }~M(B_c) =6275(7)~{\rm MeV}.
\label{3.4}
\ee

The masses of the triplet and singlet HL mesons are calculated taking
into account the HF interaction  which in general contains both
perturbative and NP contributions:
\be
 \Delta_{\rm HF} (nS) =\Delta_{\rm HF}^{\rm P} +\Delta_{\rm HF}^{\rm NP},
\label{3.5}
\ee
where the P term with the one-loop correction is taken from
\cite{42}:
\be
 \Delta_{\rm HF}^{\rm P} (nS) =\frac89
\frac{\alpha_{\rm HF}(\mu)}{\omega_1 \omega_2} |R_n(0)|^2 \left( 1+
\frac{\alpha_{\rm HF}}{\pi} \rho\right).
\label{3.6}
\ee
Here $\rho=\frac{5}{12} \beta_0 -\frac83 - \frac34 \ln 2$  and for
$n_f=3$, $\rho(n_f=3)=0.5635$, so that the one-loop correction is about
6\%. From  here on for simplicity we use the notation $\omega_1,
\omega_2$ instead of $\lan \bar \omega_1\ran_n, ~\lan\bar
\omega_2\ran_n$.

The NP contribution to the HF splitting was derived in \cite{43}
and with good accuracy it is given by the m.e.:
\be
 \Delta_{\rm HF}^{\rm NP} (nS)\approx 1.20 \frac{\pi^2}{18}
 \frac{G_2}{\omega_1\omega_2}
 \left\lan rK_1\left(\frac{r}{T_g}\right) \right\ran_{nS}.
\label{3.7}
\ee
Here $G_2$ is the gluonic condensate, for which we take the value
$G_2=0.043$ GeV$^4$ \cite{43}, which  provides the correct value
of the string tension $\sigma=0.18 $ GeV$^2$. For the $D$ and
$D_s$ mesons the NP term (\ref{3.7}) appears to be not small (as
well as  for the $J/\psi-\eta_c$ splitting):  around 10 MeV if the
gluonic (vacuum) correlation length is $T_g=0.2$ fm \cite{44}.
However, at present the value of $T_g$ is not  known with high
accuracy and $T_g=0.28$ fm was obtained from unquenched lattice
data \cite{45}. Due to this uncertainty in $T_g$ we have two
possibilities to describe the HF effects in HL mesons:

A. In the case of $T_g\approx 0.2$ fm, $\Delta_{\rm HF}^{\rm
NP}\approx 10$ MeV is not large. Then in the perturbative term,
Eq.~(\ref{3.6}), we need to take a rather large coupling:
$\alpha_{\rm HF} =0.40$ for the $D$ and $D_s$ mesons and
$\alpha_{\rm HF}=0.32$ for the $B$ and $B_s$ mesons, to reach
agreement with experiment.

B. In the case of large gluonic length, $T_g=0.3$ fm, the NP contribution
$\Delta^{\rm NP}_{\rm HF}$ appears to be larger, about 22 MeV, and
therefore the coupling $\alpha_{\rm HF}$ can be taken smaller. In this
case for the $D$ and $D_s$ mesons the value $\alpha_{\rm HF} =0.365$
(rather close to that for the  $J/\Psi -\eta_c$ splitting \cite{46})
gives rise to agreement with the experimental HF splitting. In
Table~\ref{tab.3} the calculated HF splittings both for $T_g=0.2$ fm and
$T_g =0.3$ fm are given.

\begin{table}
\caption{The hyperfine splittings (in MeV) in HL mesons. Case A:
$T_g=0.20$ fm, $\alpha_{\rm HF} =0.40$ for the $D$ and $D_s$ mesons and
$\alpha_{\rm HF} =0.32$ for the $B$ and $B_s$ mesons. Case B: $T_g
=0.30$ fm, $\alpha_{\rm HF}=0.365$ for the $D$ and $D_s$ mesons and
$\alpha_{\rm HF} = 0.305$ for the  $B$ and $B_s$ mesons.\label{tab.3}}

\begin{center}

\begin{tabular}{|l|l|l|l|l| }\hline
&\multicolumn{4}{|c|}{$T_g =0.20$ fm}\\\hline
~Multiplet &~$\Delta^{\rm P}_{\rm HF}$~~ &~$\Delta_{\rm HF}^{\rm NP}$~
 &~$\Delta_{\rm HF}$ (tot)~&~$\Delta_{\rm HF}$ (exp) \\
\hline
$D^{*} - D$     &~131.6~&~8.9~&~140.5~&~140.64$\pm$0.10~\\
$D_s^{*} - D_s$ &~130.7~&~8.9~&~139.6~&~143.8$\pm$0.4~\\
$B^{*} - B$     &~43.5~&~1.5&~45.0~&~$45.78\pm0.35$~\\
$B_s^* - B_s$   &~44.2~&~1.4&~45.8~&~44.2$\pm$1.8~\\
\hline
 &\multicolumn{4}{|c|}{$T_g =0.30$ fm}\\\hline
$D^{*} - D$     &~119.4~&~21.8~&~141.2~&~140.6$\pm$0.1~ \\
$D_s^{*} - D_s$ &~118.6~&~21.6~&~140.2~&~143.8$\pm$0.4~\\
$B^{*} - B$     &~41.3&~3.7~&~45.0~&~$45.8\pm0.4$~\\
$B_s - B$       &~42.0~&~3.4~&~45.4~&~44.2$\pm$1.8~\\
\hline
\end{tabular}
\end{center}

\end{table}

Taking the HF splittings from Table 3 the masses of the singlet and
triplet states  can be  calculated; they are presented in
Table~\ref{tab.4} ($M_{\rm cog} (1S)$ are taken from  Table~\ref{tab.2}).

\begin{table}
\caption{ The masses (in MeV) of the  ground states (for $T_g
=0.3$ fm).\label{tab.4}}
\begin{center}
\begin{tabular}{|l|l|l|l|l|l|l|l|l| }\hline
& $D^\pm$ &$D^{*\pm}$&$D_s$& $D_s^*$& $B$& $B^*$ &$B_s$&
$B_s^*$\\
\hline
this paper & 1869.1&2010.3&1966.8&2107.1& 5279.3&
5324.2& 5362.0&5407.4\\
\hline Exper.$^{a)}$& 1869.3& 2010.0& 1968.2& 2112.0&5279.0
&5325.0& 5367.7&5411.7\\
&$\pm 0.4$&$\pm 0.4$&$\pm 0.4$&$\pm 0.6$&$\pm 0.5$&$\pm 0.6$&$\pm
1.8$&$\pm 3.2$\\
\hline
\end{tabular}

$^{a)}$  The experimental numbers are taken from PDG [16] and
$M(B^*_s)$ from Ref.~\cite{19}.
\end{center}
\end{table}

The calculated triplet and singlet masses for the ground states
appear to be in good agreement with the experimental numbers (with
an accuracy $\la 5$ MeV). In our analysis we have observed that
for the $D$ and $D_s$ mesons the value $M_{\rm cog} (1S)$ is very
sensitive to the pole mass of the $c$ quark and high accuracy can
be reached only for $m_c$ (pole)$= 1.40$ GeV; at the same time we
cannot exclude that for the $B_c$ meson the choice of $m_c=1.38$
GeV is also possible.

For the $B_c$ meson, with the use of $\alpha_{\rm HF}=0.26$ and $M_{\rm
cog} =6313$ MeV from Table~\ref{tab.2}, we obtain $\Delta_{\rm HF}
(B_c) =44$ MeV, which  gives $M(B_c)=6280$ MeV  close to $M(B_c)(\rm
exp) =6275(7)$ MeV \cite{42} and $M(B_c^*)=6324$ MeV.

Finally, in Table~\ref{tab.5} we give the calculated masses
$M(2S)$ for the first  radial excitations, which are not yet found
in experiment. Our prediction for the singlet and triplet masses
of the radially excited HL mesons  strongly depends on the value
of $\alpha_{\rm HF}$ taken. If for the $D(2S)$ and $ D_s(2S)$
mesons one takes the value $\alpha_{\rm HF} (\mu_2) =0.30$, as in
the case of $\eta_c(2S) $ \cite{46}, then the values  $M(2^1S_0)$
and $M(2^3S_1)$,  given in Table~\ref{tab.5}, are obtained.

\begin{table}
\caption{ The  masses $M_{\rm cog}(2S)$, $M(2^1S_0)$ and $M(2^3S_1)$
(in MeV) for heavy-light mesons ($\alpha_{\rm HF} =0.30)$. \label{tab.5}}
\begin{center}
\begin{tabular}{|l|l|l|l|l|l|l| }\hline
 & & $D(2S)$ &$D_s(2S)$&$B(2S)$& $B_s(2S)$& $B_c(2S)^{a)}$\\\hline
~this paper~&$M_0(2S)$ &~2758&~2797&~5998&~6034&~6868\\
 &$M_{\rm cog}(2S)$    &~2615&~2702&~5888&~5966&~6852\\
& $M(2^1S_0)$          &~2560&~2646&~5864&~5941&~6821\\
 &  $M(2^3S_1)$        &~2633&~2721&~5896&~5974&~6862\\
\hline
~from Ref.~\cite{4}~&$M(2^1S_0)$&~2589&~2700&~5886&~5985&\\
 & $M(2^3S_1)$                  &~2692&~2806&~5920&~6019&\\
\hline
~from Ref.~\cite{2}~&$M(2^1S_0)$ &~2580&~2670&~5900&~5980&~6855\\
 &$M(2^3S_1)$                   &~2640&~2730&~5930&~6010&~6887\\
\hline

\end{tabular}

\end{center}
${a)}$ For the $B_c(2S)$ meson we use the value $\alpha_{\rm HF} =0.26$.
\end{table}

The triplet and singlet masses, calculated here, are rather close
to those from \cite{2,4}, nevertheless for the $2^1S_0$ states our
numbers  are systematically lower by $\sim 30 - 50$ MeV.

 In this paper we do not consider orbital excitations of HL mesons, it
will be done in our next paper. Still, we would like to notice
that our values of $M(B_{s1})$ and $M(B_{s2})$ lie in the region
$5.82-5.83$, i.e., approximately  by 100 MeV lower than in
Ref.~\cite{4} and close to the recent experimental data \cite{19}.

\section{Decay constants of heavy-light mesons}
\label{sect.4}

The general formula for the decay constants $f^2_\Gamma$ (23) can be
rewritten (later on we shall use the notation $\omega_1,\omega_2$ for
$\lan\omega_1\ran,\lan\omega_2\ran$ )
as follows
\be
 (f^n_\Gamma)^2(nS)= \frac{3\lan Y_\Gamma\ran}{2\pi \omega_1
 \omega_2M_n} |R_n(0)|^2,
\label{4.1}
\ee
which contains the w.f. at the origin $R_n(0)
=\varphi_n(0)/\sqrt{4\pi}$, the average kinetic energies
$\omega_1, \omega_2$, the meson mass $M_n=M(nS)$, and also the
m.e.  $\lan \vep^2\ran$ in the factor $\lan Y_\Gamma\ran $. All
these auxiliary values are given in Appendix~\ref{app.D}
(Tables~\ref{tab.9} and \ref{tab.10}). Then the values of $f_{\rm
P}$ can be easily calculated. They are given in Table~\ref{tab.6},
together with the experimental, unquenched  and quenched lattice
data, and some other theoretical analyses.

\begin{table}
\caption{The decay constants $f_{\rm P}$ (in MeV).\label{tab.6}}
\begin{center}
\begin{tabular}{|l|l|l|l|l|l| }
\hline
 &~$D$ &~$D_s$&~$B$& $~B_s$&~ $B_c$\\
\hline
~Ref.~\cite{20}&~206&~252&~174&~$-$&~~$-$\\
~Ref.~\cite{7} &~230(25)&~248(27)&~196(29)&~216(32)&~322(42)\\
~Ref.~\cite{6} &~234&~268&~189&~218&~~$-$\\
\hline
~lattice \cite{11,14}&~235(22) &~266(28)&~-&~206(10)&~-\\
~quenched &&&&&\\
\hline
~lattice \cite{11,13} &~201(20)&~249(19)&~216(38)&~259(32)&~440(2)\\
~$n_f=2+1$&&&&&\\
\hline
~this paper        &~210(10)&~260(10)&~182(8)&~216(8)&~438(10)\\
\hline
~experiment &~222.6(20)$^{a)}$&~280(23)$^{a)}$&~$160^{+50}_{-80}$ $^{b)}$&&\\
            &                 &               &~229(70)$^{c)}$& &\\
\hline

\end{tabular}

$^{a)}$ The experimental values are taken from Ref.~\cite{17}.

$^{b)}$ BaBar data \cite{18}

$^{c)}$ Belle data \cite{18}

\end{center}

\end{table}

From Table~\ref{tab.6} one can see that our central value of
$f_{B_d}$ is 15\% smaller than the one in unquenched lattice QCD
\cite{13}, but rather close to $f_B$ in relativistic models
\cite{6,7}.

For the analysis of experimental data on direct measurements of the
leptonic decay, $P\to l\nu$, it is important to know the ratios of the
decay constants, which in our calculations  are,
\be \frac{f_{D_s}}{f_D} =1.24(3),\quad \frac{f_{B_s}}{f_B}
=1.19(3), \quad \frac{f_{D_s}}{f_{B_s}}=1.20(3);
\quad\frac{f_D}{f_B}=1.15(2). \label{52a} \ee
These ratios  are in good agreement with recent lattice data
(unquenched) \cite{13,15} and close to the experimental number
obtained by the CLEO collaboration  with $f_{D_s}/f_D =1.27(14)$
\cite{17}. It is of interest to compare these ratios with other
theoretical calculations, which are typically smaller than our
numbers, and  also
with recent lattica data  (see Table~\ref{tab.7}).\\

\begin{table}
\caption{The  ratios $f_{D_s}/f_{D},~~ f_{B_s}/f_B$, and
$f_{D_s}/f_{B_s}$.\label{tab.7}}
\begin{center}
\begin{tabular}{|l|l|l|l| }
\hline
 &~$f_{D_s}/f_D$&~$f_{B_s}/f_{B}$&~${f_{D_s}/f_{B_s}}^{a)}$ \\
\hline
~RPM \cite{6} &~1.15&~1.15&~1.23\\
\hline
~BS \cite{7}  &~1.08(1) &~1.10(1)&~1.15(1)\\
\hline
~lattice \cite{13}&~1.24(8)&~1.20(4)&~1.01(8) \\
unquenched &&&\\
\hline
~this work&~1.24(3)&~1.19(3)&~1.20(3)\\
\hline
~experiment&~1.27(14)& &\\
\hline
\end{tabular}

$^{a)}$ The ratio of the central values is  taken for $f_{D_s}$
and $f_{B_s}$.
\end{center}

\end{table}

Due to the large theoretical errors in the ratios $ \zeta_D
=f_{D_s}/f_D$ and $\zeta_B= f_{B_s}/f_B$ (see Table~\ref{tab.7})
one cannot judge what is the true value of $\zeta_D$ and
$\zeta_B$, however, from our general formula (\ref{4.1}) for
$f_{\rm P}$ it follows that $f_{D_s}$ and $f_D$ as well as
$f_{B_s}$ and $f_{B}$ have to differ by $\sim 20\%$. This happens
due to the presence in Eq.~(\ref{4.1}), through $\langle
Y_\Gamma\rangle$, of a term proportional to $m_1m_2$: $m_1=0.008$
GeV for $D^+$ and $m_1 =0.18$ GeV  for $D^+_s$, $ m_2 =1.40$ in
both cases.  If one neglects this term then the value $f_{D_d}
=f_{D_u}=208$ MeV  practically does not change, while for the
$D_s$ meson one obtains the essentially smaller number 215 MeV,
instead of $f_{D_s}=260$ MeV in the case with $m_1 =m_s=0.18$ GeV.
Thus, the ratios $\zeta_D$ and $\zeta_B$ appear to be very
sensitive to what one takes for the pole mass of the $s$ quark and
can be used as a convenient criterium to choose $m_s$ at low
renormalization scale. Note that the factor
$|R_1(0)|^2/(\omega_1\omega_2)=0.35(1)$ turns out to be the same
both for the $D$ and $D_s$ mesons. Our number for
$\zeta_D=1.24(3)$ (with $m_s=0.18$ GeV) is in agreement with the
experimental number $\zeta_D(\exp)=1.27(14)$ \cite{17}.

The first radial excitations of HL mesons are considered here
neglecting open channels which can decrease the w.f. at the origin. In
this approximation
\begin{eqnarray}
 & & f_{\rm P}(D(2S))=167 {\rm~ MeV}, \;
 f_{\rm P}(D_s(2S))=201~{\rm MeV},
\nonumber \\
 & &
 f_{\rm P}(B(2S))=168 {\rm~ MeV},\; f_{\rm P}(B_s(2S))=194 {\rm~ MeV},~\;
 f_{\rm P}(B_c(2S))=347{\rm~ MeV}.
\label{53a}
\end{eqnarray}
From these numbers it follows that for the $2S$ states
$f_{D_s}/f_{D}=1.20$ and  $f_{B_s}/f_{B}=1.15$ are changed only by
about $\approx 4\%$, although the  absolute  values of $f_{\rm
P}(2S)$ appear to be $\approx 20\%$ smaller than for the ground
states.

One can compare our results for $f_{\rm P}$ with other theoretical
calculations (see Table~\ref{tab.6}). The agreement with
potential-model results \cite{6,7} is evident with the only
exception for $f_{\rm P}(B_c)$: our number $f_{\rm P}(B_c)=439$
MeV is about 30\% higher, and is close to the one calculated in
unquenched lattice QCD, where $f_{\rm P}(B_c)$ (lattice) = 420
(20) MeV in \cite{13}. A detailed analysis of $f_{B_c}$ (in the
framework of the potential model and QCD sum rule approach with
radiative corrections taken into account) \cite{47} gives the
value $f_{B_c}\approx 400$ MeV, which is approximately $10\%$
lower than our number.

Finally, some remarks about decay constants in V channels. As seen
from the expressions for $\lan Y_{\rm V}\ran$ and $\lan Y_{\rm P}\ran$,
$f^2_{\rm V}$ has to be larger than $f^2_{\rm P}$ (since $M_{\rm V}$ and $M_{\rm
P}$ are close to each other), while $\lan \vep^2\ran$ enters with
different signs
\be
 \frac{f_{\rm V}^2(nS)}{f^2_{\rm P}(nS)} =
 \frac{(m_1m_2+\omega_1\omega_2 + \frac13\lan \vep^2\ran_{nS})}
 {(m_1m_2+\omega_1\omega_2 -\lan\vep^2\ran_{nS})}
 \frac{M_{\rm P}(nS)}{M_{\rm V}(nS)} >1.
\label{54a}
\ee
These ratios (for the ground states) are given in Table~\ref{tab.8}.
For the heavy $B_c$ meson this ratio is approaching unity.

\begin{table}
\caption{The vector decay constants $f_{\rm V}$ (in MeV) and the ratio
$f_{\rm V}/f_{\rm P}$ for the $1S$ states in heavy-light
mesons.\label{tab.8}}
\begin{center}
\begin{tabular}{|l|l|l|l|l|l| }
\hline
&~$D^*$ &$~D^*_s$&~$B^*$& ~$B^*_s$& ~$B^*_c$\\
\hline
 ~$\lan \vep^2\ran$ in GeV$^2$&~0.273&~0.291&~0.359&~0.383&~0.784\\
\hline
 ~$f_{\rm V}$ [5]& ~340(23)&~375(24)&~238(18)&~272(20)&~418(24)\\
\hline
 ~$f_{\rm V}$, this work& ~273(13)& ~307(18)& ~200(10)&~230(12) &
~453(20)\\
\hline
 ~$f_{\rm V}/f_{\rm P}$ \cite{5} &~1.48(26)& ~1.51(26)& ~1.21(27)&
~1.26(28)&~1.30(24)\\
\hline
 ~$f_{\rm V}/f_{\rm P}$~this paper &~1.27(5)&~1.17(4)&~1.08(4)&~1.07(4)&~1.03(3)\\

\hline

\end{tabular}

\end{center}

\end{table}

From Table~\ref{tab.8} one can see that our numbers for the vector
decay constants are systematically lower than $f_{\rm V}$ (central values)
from Ref. \cite{5}: by $\approx  20\%$ for the $D$ and $D_s$ mesons and
by $\approx 15\%$ for the $B$ and $B_s$ mesons, although their values
lie within the large theoretical errors.

\section{Approximations}
\label{sect.5}

We discuss here the approximations we made and the accuracy of our
results. The starting expression for the current correlator (\ref{7})
is exact, because the FFSR is an exact representation of the meson
Green's function. The main approximation refers to the transition from
the FFSR path-integral to the local Hamiltonian formalism (\ref{18})
neglecting quark pair and hybrid production and taking spin-dependent
interactions as a perturbation. The accuracy of this approximation is
determined by several factors:

1) Neglect  of hybrid excitations in the Wilson loop, which
actually leads to a multichannel Hamiltonian. The hybrid admixture
was shown to be small, of  the order of few percent for ground
state mesons \cite{48,49}.

2) The use of the EA to define the w.f. at the origin and some
m.e. The corresponding accuracy was checked in \cite{37} and shown
to be around 5\%.

One should stress that the SH contains only well defined
fundamental parameters: the pole quark masses, $\Lambda_{\rm
QCD}$, and the universal string tension $\sigma$, and does not
contain any fitting parameters, in particular, there is no overall
constant often used in the static potential, or in the mass. The
accuracy of the SH was checked for light mesons \cite{24}, heavy
quarkonia \cite{25}, heavy-light mesons \cite{26}, glueballs
\cite{50}, and hybrids \cite{49}. In all cases  meson masses are
in agreement with experimental and lattice data with an accuracy
of a few percent.

Concerning the decay constants $f_{\rm P}$, a special sensitivity
occurs in the w.f. at the origin $|\varphi_n(0)|^2$, i.e., to the
behaviour of the GE potential (or the vector coupling $\alpha_{\rm
st}(r)$) at small distances, not only in the AF region $(r\la 0.1$
fm) but also in the region $0.1$ fm $\la r \la 0.3$ fm, which  in
\cite{39} called the intermediate region.  This behavior  is known
quite well for the perturbative part, where $\Lambda_{\rm QCD} $
is well known (for $n_f=5$ $\Lambda^{(5)}_{\overline{MS}}
=217^{+25}_{-23}$ MeV \cite{16}). The major uncertainty comes from
two sources:

(i) the behaviour of the
spin-dependent part, in particular, the HF interaction, where smearing
of the $\delta $ function can  drastically change the wave function for
systems of small size, $R\la 0.4$ fm. For HL mesons with $R\ga 0.6$ fm
this effect is becoming less important (for more discussion of the
influence of the HF interaction see \cite{51}).

(ii) We  estimate the
accuracy of the resulting $|\varphi_n(0)|^2$ better than $\pm 5\%$. Thus
the accuracy of our computed values of $f_{\rm P}$  is expected to be
$\la 8\%$, while the ratio of decay constant has better accuracy, $\la
4\%$.

At this point one should discuss the effects which were unaccounted for
till now. First of all, this concerns the radiative corrections due
to transverse gluon exchanges and higher loops. These corrections
contain the (pseudo) evolution factor, governed by the operator
anomalous dimension, and considered in \cite{52} two decades ago,
\be
 X_M=\left( \frac{\alpha_s(m_b)}{\alpha_s(m_c)}\right)^{-\frac{6}{33-2n_f}}
 \equiv x^{\frac{6}{33-2n_f}}.
\label{58}
\ee
One can estimate that when going from $B$ to $D$ mesons this factor
changes only by 4\%.

Nowadays the radiative corrections are done within HQET and known
to three-loop accuracy \cite{53}, e.g. for $f_D$ one has
\be
 f_D=\frac{1}{\sqrt{m_c}}\left[1+c_1\frac{\alpha_s^{(4)}(m_c)}{4\pi} +
 c_2^{(3)} \left(\frac{\alpha_s^{(4)}(m_c)}{4\pi}\right)^2+...\right]
 F^{(3)}(m_c)+O\left(\frac{\Lambda_{\rm QCD}}{m_c}\right),
\label{59}
\ee
where $F^{(3)}(m_c)$ is subject to the operator anomalous dimension
correlations, and $c_1=-2C_f$ etc. As a result, e.g. for the ratio
$f_B/f_D$, one has \cite{53}
\be
 \frac{f_B}{f_D} =\sqrt{\frac{m_c}{m_b}} X_M\left[1+r_1(x-1)
 \frac{\alpha_s^{(4)}(m_b)}{4\pi} + O\left(
 \left(\frac{\alpha_s^{(4)}}{4\pi}\right)^2\right)\right]+
 O\left(\alpha^3_s,
 \frac{\Lambda_{\rm QCD}}{m_{c,b}}\right)
\label{60}
\ee
with
$r_1=\frac{56}{75} \zeta_2 +\frac{4403}{1875}\cong3.58$, and
$x\approx 1.56$. One can see that loop corrections contribute less
than 4\% and can be neglected within the accuracy of HQET and our
approximations.

From Eq.~(\ref{60}) it follows that in HQET the ratio $f_B/f_D<1$  as
it happens in relativistic models \cite{2,5,6,7} and also in our
calculations where $f_B/f_D\approx 0.87(2)$, while in unquenched
lattice data this ratio is larger than unity, but has large
computational error ($\sim 20\%)$.

We now turn again to Table~\ref{tab.6} and discuss our results in
comparison to other calculations. The first important point, which
should be stressed, is that our input is minimal and fundamental, e.g.
the pole masses in Table~\ref{tab.1} correspond to the current masses
quoted by PDG \cite{16}. This is in contrast to many relativistic
potential models with spinless  Salpeter \cite{2} or Bethe-Salpeter
Hamiltonian \cite{5,7}, where the constituent quark masses are used as
input. In addition an overall constant is usually introduced in the
interaction. In many respects our approach can be considered on the
same grounds as lattice simulations, or the QCD-sum-rule approach
(where instead of $\sigma$ several additional condensates are used).
Moreover, the advantage of our approach is that excited states can be
considered as well as the ground states. For excited states a quark
(antiquark) of a given $nL$ state has its characteristic
``constituent'' mass $\lan \bar \omega\ran_{nL}$, which grows for
higher $nL$.  Due to this effect, and also to the negative string
correction, in our approach the masses of the radial and orbital
excitations appear to be smaller ($\sim 20 - 40$ MeV) than in other
relativistic models (see Table~\ref{tab.5}).

Looking at Table~\ref{tab.6} one notices a good agreement (within
$6-10$\%) of our results with unquenched lattice data and experiment
(for $f_D$ and $f_{D_s}$). Thus one can conclude that the Field
Correlator Method and its essential part, the effective String
Hamiltonian, appears to be successful in the prediction of $f_\Gamma$
as  well as in other tests done so far \cite{21,22,23,24,25,26}.

\section{Conclusions}
\label{sect.6}

In this paper we have studied the current correlator $G_\Gamma(x) =\lan
j_\Gamma (x) j_\Gamma (0) \ran $ and the integral $J_\Gamma = \int
G_\Gamma (x) d\vex$ in an arbitrary channel $\Gamma$ with the use of
the FFS path-integral representation. This method allows one to express
decay constants $f_\Gamma (nS)$ for HL mesons through well-defined
characteristics of the relativistic SH  which was successfully used
before in light mesons and heavy quarkonia.

It is essential that the SH does not contain any fitting parameters, being
fully defined by universal fundamental values: the conventional pole
masses, the string tension $\sigma$, and the strong vector coupling
$\alpha_{\rm st} (r)$.

The analytical expressions, obtained here for $f_{\rm P}$, show that
the decay constants $f_{D_s}$ and $f_{B_s}$ for the ground states
have to be  always larger
by 20-25\% than $f_D$ and $f_B$ because of the large difference between
the pole (current) masses of the strange  and light $u(d)$ quarks. This
theoretical statement is supported by recent experimental data
\cite{17}.

In our analysis we have observed that
\begin{itemize}

\item The calculated masses $M(1^1S_0)$  and $M(1^3S_1)$ of all HL
mesons agree with the experimental numbers within $\pm 5$ MeV.
Our prediction for $B_c^*$ is $M(B^*_c)=6325(10) $ MeV.

\item The calculated masses of the first radial excitations $M(2^1S_0)$
for the $D$ and $D_s$ mesons appear to be $\approx 40$ MeV lower and
for the $B$, $B_s$, and $B_c$ mesons $\approx 20 - 30$ MeV lower than
the numbers from  Refs.~\cite{2,4}.

\item For the decay constants the values $f_D=210(10)$ MeV and $f_{D_s}
=260(10)$ MeV are obtained. Their ratio $f_{D_s}/f_D=1.24(3)$ is close
to the experimental number 1.27(14) \cite{17}.

\item Our decay constants $f_B=182(8)$ MeV and $f_{B_s}=216( 8)$ MeV
give the ratio $f_{B_s}/f_B=1.19(3)$ which agrees with a recent
unquenched lattice number \cite{11,13}.

\item In the V channel the ratio $f_{\rm V}/f_{\rm P}$ is
monotonically decreasing while going from the $D$ meson to the
heavier mesons: it is equal to 1.27(6), 1.17(4), 1.08(4),
1.07(4), 1.03(3) for the $D$, $D_s$, $B$, $B_s$, and $B_c$ mesons,
respectively. For the $D_s^*$ and $B_s^*$ mesons calculated here,
$f_{\rm V}$ turned out to be $\approx 20\%$ smaller than in Ref.~\cite{5}.
\end{itemize}

\appendix

\section{Free quark propagator in the einbein path-integral representation}
\label{app.A}

\setcounter{equation}{0} \def\theequation{A.\arabic{equation}}

One starts with the FFSR for the free quark propagator, which can be
written as
\be
 S(x,y) = (m-\hat \partial) \int^\infty_0 ds (Dz)_{xy} \exp (-K),
\label{A1.1}
\ee
and introduces the einbein
variable, or the dynamical mass, $\omega(t)$ as in (\ref{3}), such
that the function  $K$ can be rewritten as
\begin{eqnarray}
 K & = & m^2s+\frac14 \int^s_0 \left(\frac{dz_\mu(\tau)}{d\tau}\right)^2d\tau
\nonumber \\
 & = & \int^T_0 dt
\left\{ \frac{m^2}{2\omega(t)}+\frac{\omega(t)}{2} +\frac{
\omega(t)}{2} \left( \frac{dz_i(t)}{dt} \right)^2 \right\}.
\label{A1.2}
\end {eqnarray}
In $(Dz)_{xy},$ Eq.~(\ref{eq.2ab}), there is an integration over the  time
components of the path, namely,
\be
 (Dz_4)\equiv \prod_k \frac{d\Delta z_4(k)}{(4\pi\varepsilon)^{1/2}}
 \delta \left(\sum\Delta z_4-T\right),
\label{A1.3}
\ee
where $T\equiv x_4-y_4$. With the use of (\ref{3}) one can rewrite the
integration element in (\ref{A1.3}) as follows $(t\equiv z_4)$,
\be
 \frac{d\Delta z_4(k)}{\sqrt{4\pi \varepsilon}} = 2d\omega(k)
 \sqrt{\frac{\varepsilon}{4\pi}} =
 \frac{d\omega(k) \sqrt{\Delta t}}{\sqrt{2\pi\omega(k)}},\;
 \sqrt{\varepsilon}= \sqrt{\frac{\Delta t}{2\omega(k)}}.
\label{A1.4}
\ee
Moreover, the $\delta$-function  in (\ref{A1.3}) acquires the form
\be
 \delta \left(\sum \Delta z_4-T\right ) = \delta ( 2\bar \omega s-T),
\label{A1.5}
\ee
where we have defined
\be
 \bar \omega=\frac{1}{s} \int^s_0 \omega(\tau) d\tau.
\label{A1.6}
\ee
As a result in (\ref{A1.1}) one can integrate  over $ds$  using the
$\delta$-function (\ref{A1.5}), and rewrite $ds (Dz)_{xy}$ as it is
shown in Eq.~(\ref{4}) of the main text.

Then one can write the Green's function as follows:
\be
 S(x,y) =(m-\hat \partial) \int \prod \frac{d^3\Delta z_i(k)}{l^3} e^{-K}
 \frac{d\omega (k)}{l_\omega(k)} \frac{d^3p}{(2\pi)^3}
 e^{i\vep\cdot(\vex-\vey-\sum\Delta \vez(k))},
\label{A1.7}
\ee
where $K$ is given in (\ref{A1.2}), and $l, l_\mu$ in Eq.~(\ref{5}).

The integration over $d^3\Delta z_i(k)$ yields
\be S(x,y) =
(m-\hat \partial) \int \frac{d^3p}{(2\pi)^3} e^{i\vep(\vex-\vey)
-\frac12 \int^T_0 dt \omega(t)  \left( 1 +
\frac{\vep^2+m^2}{\omega^2(t)} \right)}\frac{1}{2\bar \omega
}(D\omega).
\label{A1.8}
\ee
Taking into account the relation,
\be
 \int^\infty_0 \frac{d\omega(k)}{\sqrt{\omega(k)}}e^{-\frac{\Delta t}{2}
 \left(\omega(k) + \frac{\vep^2+m^2}{\omega(k)}\right) }
 =\sqrt{\frac{2\pi}{\Delta t}} e^{-\Delta t \sqrt{\vep^2+m^2}},
\label{A1.9}
\ee
one has for the scalar part $G(x,y)$, defined by $S=(m-\hat \partial) G$,
the following expression:
\be
 G(x,y) =\int \frac{d^3p}{(2\pi)^3} \frac{e^{i\vep\cdot (\vex-\vey)
 - \int^T_0 dt \sqrt{\vep^2+m^2}}}{2\sqrt{\vep^2 +m^2}},
\label{A1.10}
\ee
where we have used  the relation following
from the stationary point in the integral (\ref{A1.9}):
\be
 \bar \omega = \frac{1}{s}\int^s_0 \omega(\tau) d\tau
 =\frac{1}{N}\sum^N_{k=1}\omega(k)=\sqrt{\vep^2+m^2}.
\label{A1.11}
\ee
The expression (\ref{A1.10}) can be compared with the conventional
integral,
\be
 G(\ver, T) =\int \frac{d^4p}{(2\pi)^4}
 \frac{e^{i\vep\cdot \ver + ip_4 T}}{p^2_4+\vep^2 +m^2}, \quad
 \ver = \vex-\vey,
\label{A1.12}
\ee
which reduces to (\ref{A1.10}) after integrating over $dp_4$ for $T>0$.
The expression  (\ref{A1.12}) is the standard form of the free
propagator.

\section{Calculation of  the factor $\lan Y_\Gamma\ran $ }
\label{app.B}

\setcounter{equation}{0} \def\theequation{B.\arabic{equation}}

Here we use and extend Appendix~1 of Ref.~\cite{20},
considering the  quark propagator (\ref{1}) in the gluonic field
$A_\mu$: $S(x,y) \equiv [m-\gamma_\mu \left(
\partial/\partial x_\mu -ig A_\mu\right)]\, G(x,y)$. The
derivative $\partial/\partial x_\mu$, acting on $G(x,y)$,
differentiates only the $\delta$-function,\\
$\delta\left (x_\mu-y_\mu-\sum^N_{k=1} \Delta z_\mu\right)$ and can be
rewritten as a derivative in $\Delta z_\mu(n)$:
\be
 \frac{\partial}{\partial x_\mu} \delta\left (x-y-\sum^N_{k=1}
 \Delta z(k)\right)=-\frac{\partial}{\partial\Delta z_\mu (N)}
 \delta\left (x-y-\sum^N_{k=1} \Delta z(k)\right) .
\label{A2.1}
\ee
Integrating by parts in the expression for $G(x,y)$, one obtains
\be
 D_\mu G(x,y) =\int^\infty_0 ds e^{-ms} (Dz)_{xy} e^{-K} \Phi_\sigma(x,y)
 \left(-\frac{\Delta z_\mu(N)}{2\varepsilon} +O(\sqrt{\varepsilon})\right).
\label{A2.2}
\ee
In the limit
$\varepsilon\to 0, \; N\to \infty$ one has
$\left.\frac{\Delta z_\mu(k)}{2\varepsilon}\to
 \frac{d z_\mu(\tau)}{2d\tau}\right|_{\tau=s}$
and using  the relation (\ref{3}), one obtains
\be
 D_\mu G(x,y) = -\omega (\tau=s) \frac{dz_\mu(t)}{dt} G(x,y),
\label{A2.3}
\ee
where the r.h.s. of (\ref{A2.3}) is a symbolic writing implying that
$\omega \frac{dz_\mu}{dt}$ should be under the integral in $G(x,y)$.
Finally, when $G(x,y)$ is expressed via the Hamiltonian $\hat H_\omega
$ (as in the definitions (\ref{18}) and (\ref{20})), then one
realizes that $\omega_i\dot{z}^{(i)}_\mu (t)=p_\mu^{(i)}$ and finds the
relation:
\be
 (m-\hat D) G = (m- i \hat p) G,
\label{A2.4}
\ee
which will  be used throughout the paper.

As a check one can see that (\ref{A2.4}) yields the correct form
of the free quark propagator in Euclidean space-time.

Consider now a  meson in the c.m. system and take into account
that for a quark one has
\be
 D_\mu^{(1)}\Rightarrow-\omega_1
 \dot{z}_\mu^{(1)} \Rightarrow i p^{(1)}_\mu, \quad
 D^{(2)}_\mu=\omega_2\dot{z}_\mu^{(2)} = -i p_\mu^{(2)},
\label{A2.5}
\ee
and $ p^{(1)}_4 = i\omega_1,\; p_4^{(2)}= i \omega_2,$ so that
$D^{(1)}_4\Rightarrow -\omega_1,\; D^{(2)}_4\Rightarrow \omega_2$,
while with the 3-momentum $\vep$, $p_i^{(1)} =-p_i^{(2)} =p_i$
\be D^{(1)}_i
\Rightarrow+ip_i,~~ D^{(2)}_i\Rightarrow i p_i.
\label{A2.6}
\ee
For the factor $Y_\Gamma$ (\ref{8}) one obtains
\be
 Y_\Gamma=\frac14 tr (\Gamma(m_1+\omega_1\gamma_4 -ip_k \gamma_k)
 \Gamma(m_2-\omega_2\gamma_4-ip_i\gamma_i))
\label{A2.7}
\ee
Inserting  $\Gamma=1,\, \gamma_5,\, i\gamma_\mu \gamma_5,\, \gamma_\mu$
in (\ref{A2.7}) one arrives at the expressions (\ref{34}) for
different channels.

\section{Calculation of the path integral for $G_\Gamma$, (Eq.~(\ref{7})),
in the free quark case}
\label{app.C}

\setcounter{equation}{0} \def\theequation{C.\arabic{equation}}

To find the solution in the general case  with $W_\sigma\neq 1$ and an
interaction depending only on the relative quark-antiquark coordinates,
we separate here the relative and c.m. coordinates for any
path-integral index $k$ as follows:
\begin{eqnarray}
 \Delta \vez_1 -\Delta \vez_2 & = & \veta,\quad
 \frac{\omega_1\Delta \vez_1+\omega_2 \Delta \vez_2}{\omega_1+\omega_2}=\verho,
\nonumber \\
 \omega_+ & = & \omega_1+\omega_2, \quad
 \omega_{\rm r} = \frac{\omega_1 \omega_2}{\omega_+},
\label{A3.1}
\end{eqnarray}
with the Jacobians:
\begin{eqnarray}
 d^3\Delta z_1 \; d^3 z_2 & = & d^3 \eta \; d^3 \rho,
\label{eq.A1a}
\nonumber \\
 \frac{d\omega_1 d\omega_2}{\sqrt{\omega_1\omega_2}} & = &
 \frac{d \omega_{\rm r} }{\sqrt{\omega_{\rm r}}}
 \frac{d\omega_+}{2 \sqrt{\omega_+-4 \omega_{\rm r}}}.
\label{A3.2}
\end{eqnarray}
Integrating (\ref{7}) for $W_\sigma\equiv 1$ over $\prod_k
d\verho (k)$ one obtains
\be
 G_\Gamma (x) = 4 N_c Y_\Gamma \int
 \frac{(D^3\eta)_{00}}{l^3_\eta 4\bar \omega_1 \bar \omega_2}
 \prod_k \frac{d\omega_+(k)}{\sqrt{\frac{2\pi}{\Delta t}} 2
 \sqrt{\omega_+ - 4 \omega_{\rm r}(k)}}
 \frac{ d \omega_{\rm r}(k)}{\sqrt{\omega_{\rm r}(k) \frac{2\pi}{\Delta t}}}
 \; e^{- \mathcal{F}_1}.
\label{12}
\ee
Here the notation $(D^3\eta)_{ab}$ means the initial $a$ and final
$b$ value of relative coordinate; in our case, evidently, for
considered current correlator $a=b=0$. Also in Eq.~(\ref{12}) we
use the notation $l_\eta = \left( \frac{2\pi\Delta
t}{\omega_r(\bar k)}\right)^{1/2}$. The quantity $\mathcal{F}_1$
is defined as
\be
 \mathcal{F}_1 =\sum_k \left\{ \frac{\omega_+(k) \Delta t}{2} +
 \frac{m^2\Delta t}{2 \omega_{\rm r} (k)} +
 \frac{\omega_{\rm r} (k) \eta^2 (k)}{2\Delta t} \right\}.
\label{A3.4}
\ee
Integration over $d\omega_+(k)$ can be easily done,
\be
 \prod_k\int \frac{d\omega_+(k)}{2\sqrt{\omega_+- 4 \omega_{\rm r}(k)}}
 \frac{\exp \left\{ -\sum \omega_+ (k)
 \frac{\Delta t}{2}\right\}}{\sqrt{2\pi\Delta t}}
 = \prod_k 2 e^{-\sum_k 2\omega_{\rm r}(k) \Delta t}.
\label{A3.5}
\ee
Thus at zero c.m.  momentum one finds
\be
 \int G_\Gamma (x) d^3\vex =
 \int \frac{N_cY_\Gamma}{\bar \omega_1 \bar \omega_2}
 \frac{(D^3\eta)_{00}}{l^3_\eta} \prod^N_{k=1}
 \frac{2 d\omega_{\rm r} (k)}{\sqrt{\frac{\omega_{\rm r}(k) 2\pi}{\Delta t}}}
 e^{- \mathcal{F}_2},
\label{A3.6}
\ee
where
\be
 \mathcal{F}_2 =\sum_k \left\{ 2 \omega_{\rm r} (k) \Delta t
 +\frac{m^2}{2 \omega_{\rm r} (k)} \Delta t
 +\frac{ \omega_{\rm r} (k) }{2\Delta t} \eta^2 (k)\right\}.
\label{A3.7}
\ee
At this point one can use the general relation \cite{30,36}
\be
 \int \frac{(D^3\eta)_{xy}}{l^3_\eta}
 e^{-\sum_k \left( \frac{ \omega_{\rm r} (k) \eta^2(k)}{2\Delta t} +
 \hat V(k) \Delta t\right)}= \langle x|e^{-\hat HT} |y\ran
\label{A3.8}
\ee
with
\be
 \hat H = \frac{\vep^2_\eta}{2\omega}_{\rm r} + \hat V(\eta),\quad
 \vep_\eta =\frac{1}{i} \frac{\partial}{\partial\veta}.
\label{A3.9}
\ee
Then for the free case, $\hat V(\eta) \equiv 0$, the r.h.s. of
(\ref{A3.8}) yields
\be
 \left\lan 0 \left| \exp \left(-\frac{\vep^2}{2 \omega_{\rm r}} T\right)
 \right| 0\right\ran
 = \int \frac{d^3p}{(2\pi)^3} \exp \left(-\frac{\vep^2}{2 \omega_{\rm r}} T
\right).
\label{A3.10}
\ee
Integrating  first over $d \omega_{\rm r} (k)$ and performing the
steapest descent method (stationary point analysis) one obtains
\be
\int\frac{2d\omega_{\rm r}(k)}{\sqrt{\omega_{\rm r}(k)}\sqrt{\frac{2\pi}{\Delta t}}}
 e^{-2 \omega_{\rm r} (k) \Delta t- \frac{(\vep^2+m^2)\Delta t}
{2 \omega_{\rm r} (k)} } = e^{-2\sqrt{\vep^2+m^2}{\Delta t}}
\label{A3.11}
\ee
and
\be
 \int G_\Gamma(x) d^3x= \int \frac{d^3\vep}{(2\pi)^3}
 \frac{N_cY_\Gamma}{\bar \omega_1\bar \omega_2} e^{-2T\sqrt{\vep^2+m^2}}.
\label{A3.12}
\ee
For equal current masses $m_1=m_2=m$ one has evidently $\bar \omega_1
=\bar \omega_2 =2\omega_{\rm r}$, and from (\ref{A3.11}) it follows
that for the stationary point (for any $k$)
$2\omega_{\rm r} =\sqrt{\vep^2+m^2}$.

To compare (\ref{A3.12}) with the standard Feynman amplitude for
the free quark loop, one can go to the momentum space,
\begin{eqnarray}
 G_\Gamma(\veq =0,q_4) & = & \int^\infty_{-\infty} d T e^{iq_4T}
 \int G_\Gamma(x) d^3x
\nonumber \\
 & = & \int\frac{d^3p}{(2\pi)^3} \frac{N_cY_\Gamma}{\vep^2+m^2}
 \left(
\frac{1}{2\sqrt{\vep^2+m^2}-iq_4}+
\frac{1}{2\sqrt{\vep^2+m^2}+iq_4}\right)
\nonumber \\
 & = &
 \int 4 N_cY_\Gamma
 \frac{d^3p}{(2\pi)^3\sqrt{\vep^2+m^2} [4 (\vep^2+m^2) + q^2_4]}
\nonumber \\
 & = & \int \frac{4N_cY_\Gamma d^4p}{(2\pi)^4 p^2 (p-q)^2}
\label{A3.13}
\end{eqnarray}

This result proves our sequence of equations for the free  case.  For
the case of interacting quarks, one should  use in (\ref{A3.8}) the
spectral decomposition in the infinite set of bound states, as it is
done in the main text, Eqs.~(\ref{18a},\ref{23},\ref{26b}).

\section{ The relativistic  String Hamiltonian}
\label{app.D}

\setcounter{equation}{0} \def\theequation{D.\arabic{equation}}

We start with the relativistic SH $\hat H_{\omega}$ for a meson
$q_1 \bar q_2$, taken in the most general form \cite{22}:
\be
 \hat H_\omega =\sum_{i=1,2}
 \left( \frac{\omega_i}{2}+\frac{m^2_i}{2\omega_i} \right) +
 \frac{p^2_r}{2\omega_{\rm r}} +V_{\rm GE}(r) +\int^1_0 d \beta
\left(\frac{\sigma^2 r^2}{2\nu} +\frac{\nu}{2}\right)+
\frac{\veL^2}{2r^2} \frac{1}{g(\omega_1,\omega_2, \sigma r)}.
\label{D.1}
\ee
Here $\omega_{\rm r} =\omega_1 \omega_2/(\omega_1+\omega_2)$
and $m_i (i=1,2)$ are the pole masses of a quark (antiquark) and
\be
 g(\omega_1, \omega_2, \sigma r) = \omega_1 (1-\zeta)^2 +\omega \zeta^2
 + \int^1_0 d\beta (\beta -\zeta)^2 \nu d\beta,
\label{D.2}
\ee
with
\be \zeta=\frac{\omega_1 +\frac12 \sigma r}{\omega_1 +\omega_2 +\sigma r}.
\label{D.3}
\ee
In the SH (\ref{D.1}) taken from  \cite{22} we have added the GE
potential, $V_{\rm GE} =-\frac43 \frac{\alpha_{\rm st} (r)}{r}$.
This can be done due to the property of additivity of the static
potential in QCD \cite{39,40}. $\hat H_\omega$  depends on the
variables $\omega_1$, $\omega_2$, and $\nu$, which have been shown
to be the canonical variables of the SH \cite{26}, and therefore
they can be defined from the extremum conditions. But first,
instead of the operator $p^2_r$ in (\ref{D.1}) we introduce
$\vep^2=p^2_r + \veL^2/r^2$ and present $\hat H_\omega$ as the sum
of two terms:
\be
 \hat H_\omega=H_\omega^{(0)} +H_{\rm str},
\label{D.4}
\ee
where
\be H^{(0)}_\omega =\sum_{i=1,2}
 \left( \frac{\omega_i}{2}+\frac{m^2_i}{2\omega_i} \right) +
 \frac{\vep^2}{2 \omega_{\rm r}} +V_{\rm GE}(r) +\int^1_0 d\beta
 \left(\frac{\sigma^2 r^2}{2\nu} +\frac{\nu}{2}\right),
\label{D.5}
\ee
and the ``string'' part of the SH is
\be
 H_{\rm str} =-\frac{\veL^2}{2r^2 \omega_{\rm r}}
 \left[ 1-\frac{\omega_{\rm r}}{{g}(\omega_1, \omega_2, r)}\right].
\label{D.6}
\ee
This term occurs only for the states with $L\neq0$ and can be
considered as a perturbation, because it gives corrections $\la 5\%$ to
the e.v. of the unperturbed Hamiltonian $H_\omega^{(0)}$ (\ref{D.5}).
(For light mesons this correction is becoming rather large only for the
states with $L\ga 5$). Then to determine the variables $\omega_1$,
$\omega_2$, and $\nu$ we use the following extremum conditions applied
to $H_\omega^{(0)}$:
\be
 \frac{\partial H_\omega^{(0)}}{\partial \omega_i} =0, \quad (i=1,2);\quad
 \frac{\partial H_\omega^{(0)}}{\partial \nu} =0.
\label{D.7}
\ee
From (\ref{D.7}) it follows that
\be
 \omega^2_i =\vep^2 + m^2_i,\quad \nu=\sigma r,
\label{D.8}
\ee
and therefore $H^{(0)}_\omega$ can be rewritten  as
\be
 H_\omega^{(0)} = \sum_{i=1,2}\sqrt{m^2_i+\vep^2}
 + \sigma r + V_{\rm GE}(r)\equiv T_R + V_0(r),
\label{D.9}
\ee
where $V_0(r)$ is just the same potential as in (\ref{9a}). From (\ref{D.9})
one can see that the kinetic term $T_R$ coincides with the one in the
spinless  Salpeter equation (SSE). The equation
\be
 (H^{(0)}_\omega+V_0(r))\varphi_{nL} (r)=M_0 \varphi_{nL} (r)
\label{D.10}
\ee
defines the e.v. $M_0(nL)$ and e.f. $\varphi_{nL}(r)$. There are two
essential differences between our SH $H_\omega^{(0)}$ with the kinetic
part
\be
 T_R=\sqrt{m_1^2+\vep^2}+\sqrt{m^2_2+\vep^2}
\label{D.11}
\ee
and many other papers where the SSE is used.

The first one is that in (\ref{D.11}) the quark (antiquark) mass is the
pole (current) mass and it is not considered to be a fitting parameter.
In Table~\ref{tab.1} we compare the input masses $ m_1, m_2$ used in
(\ref{D.11}) and the constituent masses from \cite{2,4,5}.

The second difference refers to the string correction $H_{\rm str}$
which can be rewritten as
\be
 H_{\rm str} =-\frac{\veL^2\sigma}{2\omega_{\rm r}}
 \left\{\frac{g_1 (\omega_1, \omega_2, r)}{r} +\frac14 \sigma
 \frac{(\omega_1-\omega_2)^2}{(\omega_1+\omega_2)^2}g_2 (\omega_1,
 \omega_2, r)\right\}.
\label{D.12}
\ee
Here
\begin{eqnarray}
 g_1 & = & \left(\frac13 -\zeta +\zeta^2\right )
 (\omega_1 +\omega_2 +\sigma r)^2 \mathcal{F}^{-1} (r,\omega_1, \omega_2),
\nonumber \\
 g_2 & = & \mathcal{F}^{-1}(r,\omega_1, \omega_2)
\label{D.14}
\end{eqnarray}
with
\begin{eqnarray}
\mathcal{F}(r_1, \omega_1, \omega_2)  & = & \omega_1
 \left(\omega_2+ \frac12 \sigma  r \right)^2
 +\omega_2 \left(\omega_1 +\frac12 \sigma r \right)^2
\nonumber \\
 & &  + \sigma r \left(\omega_1 +\omega_2 +\sigma r  \right)^2
 \left(\frac13 -\zeta +\zeta^2 \right).
\label{D.15}
\end{eqnarray}
The string correction to a meson mass, $\Delta_{\rm str} (nL) = \lan
H_{\rm str}\ran$, can be calculated with the use of the expressions
(D.12-D.14), in which $\omega_1$, $\omega_2$ can be replaced
by their averaged values with a good accuracy.

For light mesons with $m_1=m_2=0$, $\omega_1=\omega_2=\omega$,
$\sigma\lan r\ran =2\omega$, and $\zeta =\frac12$, the second term in
(\ref{D.12}) is absent and
\be
 \lan \mathcal{F} (r,\omega)\ran =\frac{32}{3} \omega^3,\;
 \lan g_1\ran =\frac{1}{8\omega};\lan g_2\ran =\frac{3}{32 \omega^3},
\label{D.16}
\ee
so that in this case ($m_1=m_2=0$) the string correction,
\be
 \Delta_{\rm str} = \lan H_{\rm str} \ran
 =- \frac{L(L+1)}{\omega}\, \frac{\sigma \lan r^{-1}\ran} {8\omega},
\label{D.17}
\ee
just coincides with the string correction obtained in \cite{24} for
light mesons. It is important that due to the negative sign of the
string correction the masses of $P$- and $D$-wave heavy-light mesons
appear to be $30 - 50$ MeV smaller in our calculations than in other
relativistic models which use the SSE equation \cite{2}.

In Table~\ref{tab.9} we give the values of the average kinetic energies
$\bar \omega_1$ and $\bar \omega_2$,  and the excitation energy
$\varepsilon (\omega_{\rm r})$,  which are needed to determine the
self-energy contribution (\ref{25a}) to $M_{\rm cog} (nS)$.

\begin{table}
\caption{ The average energies $\bar \omega_i(nS)=\left\lan
\sqrt{m^2_i+\vep^2}\right\ran_{nS}(i=1,2)$, the reduced mass
$\omega_{\rm r}$, and the excitation energy $\varepsilon_n (\omega_{\rm r})$
(in MeV) for the $1S$ and $2S$ heavy-light mesons.\label{tab.9}}
\begin{center}
\begin{tabular}{|l|l|l|l|l|l|}\hline
 Meson   &~$D$~&~$D_s$~&~$B$~&~$B_s$~&~$B_c$~\\
\hline
~$\bar \omega_1(1S)$~           &~507~&~559~&~587~&~639~&~1662~\\
~$\bar \omega_2(1S)$~           &~1509~&~1515~&~4827~&~4830~&~4869~\\
~$\omega_{\rm r}(1S)$~           &~379~&~408~&523~&~564~&~1238~\\
~$\varepsilon_1(\omega_{\rm r})$~&~541~&~534~&~432~&~406~&~149~\\
\hline
~$\bar \omega_1(2S)$~     &~643~&~692~&741~&~789~&~1732~\\
~$\bar \omega_2(2S)$~           &~1585~&~1590~&~4862~&~4865~&~4898\\
~$\omega_{\rm r} (2S)$ ~          &~457~&~482~&643~&~679~&~1279~\\
~$\varepsilon_2(\omega_{\rm r})$ &~1164~&~1124~&~985~&~959~&~687~\\
\hline

\end{tabular}

\end{center}

\end{table}

To calculate the decay constants $f_{\rm P}$ and $f_{\rm V}$ we need also  to
know the w.f. at the origin $\varphi_n(0) =R_n(0)/\sqrt{4\pi}$ and the
m.e. $\lan \vep^2\ran_{nS}$. Their values are given in
Table~\ref{tab.10}.

\begin{table}
\caption{The w.f. at the origin $|R_n(0)|^2$ (in GeV$^3$) and $\lan
\vep^2\ran_{nS} $  (in GeV$^2$) in einbein approximation for the $1S$
and $2S$ states of heavy-light mesons.\label{tab.10}}
\begin{center}
\begin{tabular}{|l|ll|ll|}\hline
 Meson   &\multicolumn{2}{c|}{$1S$}
 &\multicolumn{2}{c|}{$2S$}\\\cline{2-5}
 &~$|R_1(0)|^2$~&~$\lan \vep^2\ran_{1S}$~&~$|R_2(0)|^2$~&~$\lan \vep^2\ran_{2S}$~\\
\hline
~$D$~&~0.272~&~0.273~&~0.266~&~0.464~\\
~$D_s$~&~0.291~&~0.290~&~0.284~&~0.482~\\
~$B$~&~0.410~&~0.359~&~0.410~&~0.599~\\
~$B_s$~&~0.455~&~0.383~&~0.439~&~0.624~\\
~$B_c$~&~1.470~&~0.784~&~1.032~&~1.023~\\
\hline

\end{tabular}\\

\end{center}

\end{table}

As seen from Table~\ref{tab.9} the ``constituent'' masses $\omega_1(2S)$
and $\omega_1(1S)$ of the lighter quark $q_1$(or$\bar q_1$) differ by
$\sim 130$ MeV for the $D$ and $D_s$ mesons and $\sim 150$ MeV for the
$B$ and $B_s$ mesons. This difference in $\omega_1(nS)$ illustrates the
statement that in  relativistic  approach there does not exist
a universal constituent  mass $\omega_i$ for different $nL$ states, but
only the pole (current) quark masses may be considered as universal input.

At this point we remind that the expressions (\ref{26b},\ref{34}) for
the decay constants have been derived in the EA approximation and
therefore to have a consistent description one needs to take also
$|R_n(0)|^2$ and $\lan \vep^2\ran_{nS}$ as calculated in the EA. In this
approximation, instead of the extremum conditions (\ref{D.7}),
different conditions may be used, \cite{26}, namely,
\be
\frac{\partial M^{\rm EA}}{\partial\tilde \omega_1} =0,\quad
 \frac{\partial M^{\rm EA}}{\partial\tilde \omega_2} =0,
\label{D.17a}
\ee
where in the EA we use the notations $\tilde \omega_1$ and $\tilde
\omega_2$ instead of $\bar \omega_1$ and $ \bar \omega_2$. The mass
$M^{\rm EA} (nL)$ satisfies the equation,
\be
 H_0^{\rm EA} \tilde\varphi_{nL}(r) = M_0^{\rm EA} (nL) \tilde\varphi_{nL}(r),
\label{D.18}
\ee
where the EA Hamiltonian \cite{26} is given by,
\be
 H_0^{\rm EA} = \sum_{i=1,2} \left(\frac{\tilde \omega_i}{2}+
 \frac{m_i^2}{2\tilde\omega_i}\right) + \frac{\vep^2}{2\omega_{\rm r}}
 +V_0(r),
\label{D.19}
\ee
and has the same interaction $V_0(r)$ and the reduced mass is given by
$\tilde\omega_{\rm r}=(\tilde\omega_1
\tilde\omega_2)/(\tilde\omega_1+\tilde\omega_2)$.
Then writing
\be
 M^{\rm EA} (nL) =\sum_{i=1,2}\left(\frac{\tilde \omega_i}{2}+
 \frac{m_i^2}{2\tilde\omega_i}\right)+ \varepsilon_{nL}
 (\omega_{r}),
\label{D.20}
\ee
the excitation energy $\varepsilon (\omega_{r})$ satisfies  the
equation
\be
 \left\{ \frac{\vep^2}{2\tilde \omega_{\rm r}}+ V_0(r) \right\}
 \tilde \varphi_{nL}(r) = \varepsilon_{nL} (\omega_{\rm r})\tilde\varphi_{nL} (r).
\label{D.21}
\ee
With the use of (\ref{D.20}) the extremum conditions (D.17) reduce to
the equations:
\be
 \tilde \omega_i^2 =m_i^2-2\tilde\omega^2_r
 \frac{\partial\varepsilon}{\partial\tilde\omega_{\rm r}} \quad (i=1,2).
\label{D.22}
\ee

The derivative $\partial \varepsilon (\tilde\omega_{\rm r})/
\partial\tilde\omega_{\rm r}$ in (\ref{D.22}) is a very smooth function
of the variable  $\tilde\omega_{\rm r}(nL)$ and  can be easily
calculated. Equation (\ref{D.21}) formally coincides with the
Schr\"{o}dinger equation but differs from the physical point of view:
the masses $\tilde \omega_i (nL)(i=1,2)$ and
$\tilde\omega_{\rm r}(nL)$), being the average kinetic energy of a
quark (antiquark), are different for every $nL$ state. In this way
relativistic corrections are taken into account in Eq.~(\ref{D.21}))
through the increase of $\tilde\omega_1$ and $\tilde\omega_2$.

In the EA the values of $\tilde \omega_i$ appear to be very close to
$\bar \omega_i$, determined from the SSE (\ref{24}) or (\ref{D.10}).
Therefore one can define $\omega_i(nL)$, solving the SSE once and
calculating the m.e.
\be
 \bar \omega_i (nL) =\lan \sqrt{\vep^2+m^2}\ran_{nL},
\label{D.23}
\ee
instead of calculations of a variety of m.e. in the solutions of
(\ref{D.21}) for different $nL$ states. However, even for $\tilde
\omega_i=\omega_i$ the e.v. $M^{\rm EA}_0 (nL)$ (\ref{D.18})
slightly differs from $M_0(nL)$ for the SSE with the same quark pole mass
$m_q$. The differences between them may be called the relativistic
correction
$\delta_R$:
\be
 M_0 (nL) \equiv M_0^{\rm EA} (nL) -\delta_R,
\label{D.24}
\ee
which can be  approximately calculated,
\be
 \delta_R \approx \frac{\lan \vep^2\ran_R +m^2_1-\omega_1^2}{2\omega_1}.
\label{D.25}
\ee
In (\ref{D.25}) the index "1" refers to the lighter  quark (antiquark),
its pole mass $m_1$ and kinetic energy $\omega_1$ (\ref{D.23}). Such a
difference between $M_0$ and $M_0^{\rm EA}$ occurs due to the fact that
\be
 \sqrt{\lan \vep^2+ m_i^2\ran_{nL}} \neq\lan\sqrt{ \vep^2+ m_i^2}\ran_{nL}.
\label{D.26}
\ee
Surprisingly, for the $D$ and $D_s$ and $B$ and $B_s$ mesons
$\delta_R$ remains almost constant, being approximately equal to 70
MeV for the ground states and about 100 MeV for the 2S states.
For the heavy $B_c$ meson $\delta_R\approx 20$ MeV is essentially
smaller.

\section{\bf The self-energy correction to the meson mass}
\label{app.E}

\setcounter{equation}{0} \def\theequation{E.\arabic{equation}}

The self-energy correction to the meson mass originates from the NP
contribution to the squared quark mass $m^2_q $ as a result of the
spin interaction of a quark with NP background gluonic field
\cite{35}. It was shown  that only due to the presence of this
correction in the meson mass it is possible to obtain a linear
Regge trajectory for light mesons \cite{24}. With the use of the old
result from \cite{35} and the recent result from \cite{54} it can be
presented as
\be
 \Delta_{\rm SE} (nL) =\sum_{i=1,2}
 \left( -\frac{1.5\sigma \eta^i_f}{\pi\bar \omega_i} +
 \frac{\sigma^2}{4\bar \omega_i [m+ \bar \omega_i/ 2
 +\varepsilon (\omega_{\rm r})]^2}\right).
\label{E.1}
\ee
Here $m_i(i=1,2)$ is the pole mass of $i$-th quark (antiquark),
$\bar \omega_i$ is determined by (\ref{D.7}), $\omega_{\rm r}$ is
the reduced mass, $\varepsilon (\omega_{\rm r})$ is defined by the
solution of the equation (\ref{D.21}). The factor $\eta^i_f$
depends on the flavor of a given quark (antiquark) $q_i (\bar
q_i)$.  The first, negative term in (\ref{E.1}) was calculated in
the Simonov's paper \cite{35} while the second, positive term is
rather small and has been derived recently \cite{54}. The
analytical expression for  $\eta^i_f$ differs for a heavy quark
with mass $m_q>T_g^{-1}$ and $m_q<T^{-1}_g$, where $T_g$ is the
gluonic (vacuum) correlation length which defines the behaviour of
the bilocal vacuum correlators $D(x)$ and $D_1(x)$ \cite{21}. The
value $T_g$ has been measured in lattice QCD and in the quenched
approximation $T_g\la 0.2$fm \cite{44} while in the unquenched
case $T_g$ is larger $(T_g\approx0.30$ fm) \cite{45}.

Introducing the variable $y=m_qT_g$ (for $c,b$ quarks $y>1$ and
for $u(d)$, $s$ quarks $y<1$) $\eta_f$ is given by the following
expressions \cite{35}:
\begin{eqnarray}
 \eta_f & = & \frac{1+2y^2}{(y^2-1)^2}-\frac{3 y^2}{(y^2-1)^{5/2}}
 \arctan \sqrt{y^2-1} ~~{~\rm for }~ y\geq 1,
\label{E.2}
\nonumber \\
 \eta_f & = & \frac{1+2y^2}{(1- y^2)^2}-\frac{3 y^2}{(1- y^2)^{5/2}}
 \ln \frac{1+\sqrt{1-y^2}}{y} ~~{\rm~for }~ y< 1.
\label{E.3}
\end{eqnarray}
For $m_u,m_d\to 0$ the quantity $\eta_f$ tends to 1, while for
$m_q\to \infty$ the value $\eta_f\to 0$.
%
%
Here for $m_u=m_d\approx 0$ and $ m_b=4.78$ GeV we have
\begin{eqnarray}
\eta_f & = & 1.0,\quad {\rm for}\; u(d)\; {\rm quarks}
\nonumber \\
 \eta_b & = & 0.03,\quad {\rm for}\;b\; {\rm quarks}.
\label{E.6}
\end{eqnarray}
However, since the value of $T_g$  and the mass of the $s$ quark are
not known with a good accuracy, the value of $\eta_s$ may essentially
differ. Then from (\ref{E.3}) we obtain
\begin{eqnarray}
 \eta_s & = & 0.87~{\rm for}~y=0.20~(T_g=0.22~{\rm fm},~m_s=0.18~{\rm GeV}),
\nonumber \\
 \eta_s & = & 0.76~{\rm for}~y=0.33~(T_g=0.32~{\rm fm},~m_s=0.22~{\rm~ GeV}).
\label{E.7}
\end{eqnarray}
and also for $c$ quarks
\begin{eqnarray}
 \eta_c & = & 0.36~(T_g=0.2~{\rm fm},~m_c=1.4~{\rm  GeV}). \label{E.8}
\end{eqnarray}
In our analysis here we use the following numbers:
\be
 \eta_{u(d)} =1.0,~\eta_s =0.65,~\eta_c=0.35,~\eta_b=0.025.
\label{E.9}
\ee
The contribution of the $b$ quark to $\Delta_{\rm SE}$ (E.1)) is
about $-1$ MeV and can be neglected.

Finally, we notice that in the first term in (\ref{E.1}) we use
the number 1.5 instead of the number 2.0 derived in \cite{35}
(where $\eta_{u(d)}=0.9$ was taken, as well as in \cite{24}). This
change is made because here we do not neglect the contribution of
the correlator $D_1(x)$ as in  the first paper of \cite{35}.
Instead at $x = 0$ we use the relation $D(0)+D_1(0)
=\frac{\pi^2}{18} G_2$, where $G_2$ is the gluonic condensate (the
details are given in second paper of \cite{54}).

\begin{acknowledgments}
This work is supported by the Federal Program of the Russian Ministry
of Industry, Science, and Technology No.40.052.1.1.1112, and by the
grant of RFBR No. 06-02-17012 and State Contract No 02.445.11.7424.
\end{acknowledgments}


\begin{thebibliography}{99}

\bibitem{1} H. Krasemann, Phys. Lett. B {\bf 96}, 397 (1980);
D. Silverman and H. Yao, Phys. Rev. D {\bf 38}, 214 (1988).
%
\bibitem{2}
S. Godfrey, N. Isgur,  Phys. Rev. D {\bf 32}, 189 (1985);
S. Godfrey, Phys. Rev. D {\bf 70}, 054017 (2004) and references therein.
%
\bibitem{3} V.V. Kiselev,  A.E. Kovalsky, A.K. Likhoded,
Nucl. Phys. B {\bf 585}, 353 (2002);
S.S.Gershtein et al., Phys. Usp. {\bf 38}, 1 (1995);
Phys. Rev. {\bf D51}, 3613 (1995).
%
\bibitem{4} M. Di Pierro, E. Eichten, Phys. Rev. D {\bf 64}, 114004 (2001).
%
\bibitem{5}
Guo-Li Wang, Phys. Lett. B {\bf 633}, 492 (2006).
%
\bibitem{6}  D. Ebert, R.N. Faustov and V.O. Galkin,
Phys. Lett. B {\bf 635}, 93(2006);
Mod. Phys. Lett. A {\bf 17}, 803 (2002);
V.O. Galkin, A.Yu. Mishurov and R.N. Faustov,
Sov. J. Nucl. Phys. {\bf 53}, 1026 (1991)
[Yad. Fiz. {\bf 53}, 1676 (1991)].
%
\bibitem{7} G. Cveti\v c, C.S. Kim, G.-L. Wang, and W. Namgung,
 Phys. Lett. B {\bf 596}, 84 (2004).
%
\bibitem{8} J. He, B. Julia-Diaz and Y.B. Dong,
Eur. Phys. J A {\bf 24}, 411 (2005);
P. Maris and C.D. Roberts, Int. J. Mod. Phys. E {\bf 12}, 297 (2003);
P. Maris and P.C. Tandy,  Phys. Rev. C {\bf 60}, 055214 (1999);
Phys. Rev. C {\bf 62}, 055204 (2000);
M. Koll et al., Eur. Phys. J. A {\bf 9}, 73 (2000).
%
\bibitem{9} S.Narison, hep-ph/0202200;
Phys. Lett. B {\bf 520}, 115 (2001);
ibid {\bf 322}, 247 (1994); ibid {\bf 198}, 104 (1987).
%
\bibitem{10} A.A. Penin and M. Steinhauser, Phys. Rev. D {\bf 65}, 054006
(2002); M. Jamin, B.O. Lange,
Phys. Rev. D {\bf 65}, 056005 (2002) and reference therein.
%
\bibitem{11} For recent reviews see T. Onogi, hep-lat/0610115;


%
\bibitem{12} A. Ali Khan et al., hep-lat/0701015;
Phys. Rev. D {\bf 65}, 054505 (2002);
ibid {\bf 67}, 059901(E) (2003);
ibid {\bf 64}, 054504 (2001).
%
\bibitem{13} C. Aubin et al., Phys. Rev. Lett. {\bf 95}, 122002 (2005);
Phys. Rev. D {\bf 70}, 114501 (2004);
M. Wingate et al., Phys. Rev. Lett. {\bf 92}, 022001 (2004);
A. Gray et al., Phys. Rev. Lett. {\bf 95}, 212001 (2005).
%
\bibitem{14}
T.W. Chiu et al., Phys. Lett. B {\bf 624}, 31 (2005);
A. J\"{u}ttner and J. Rolf, Phys. Lett. B {\bf 560}, 59 (2003);
(ALPHA Collaboration) J. Rolf et al.,
Nucl. Phys. Proc. Suppl. {\bf 129}, 322 (2004); hep-lat/0309072.
%
\bibitem{15} P. Zweber, hep-ex/0701018 and references therein.
%
\bibitem{16} Particle Data Group, S. Eidelman et al.,
Phys. Lett. B {\bf 592}, 1 (2004).
%
\bibitem{17} (CLEO Collaboration) M. Artuso et al.,
hep-ex/0607074;Phys. Rev. Lett. {\bf 95}, 251801 (2005);
(CLEO Collaboaration) G. Bonvicini et al., Phys. Rev. D {\bf 70}, 112004 (2004).
%
\bibitem{18} (BELLE Collaboration) K. Ikado et al., hep-ex/0604018;
(BaBar Collaboration) B. Aubert et al., hep-ex/0608019.
%
\bibitem{19} R. Mommsen, hep-ex/0612003;
(CLEO Collaboration) G. Bonvicini et al., Phys. Rev. Lett. {\bf
96}, 022002 (2006); (CDF Collaboraton) A.Abulencia et al. Phys.
Rev. Lett. {\bf 96}, 082002 (2006); O.Aquinas et al., (CLEO
Collaboration) Phys. Rev. Lett. {\bf 96}, 152001 (2006).
%
\bibitem{20} Yu.A. Simonov, Z. Phys. C {\bf 53}, 419 (1992).
%
\bibitem{21} A. Di Giacomo, H.G. Dosch, V.I. Shevchenko, Yu.A. Simonov,
Phys. Rep. {\bf 372}, 319 (2002);
Yu.A. Simonov, Phys. Atom Nucl. {\bf 67}, 846; 1027 (2004); hep-ph/0302090,
hep-ph/0305281.
%
\bibitem{22} A.Yu. Dubin, A.B.Kaidalov, Yu.A.Simonov,
Phys. Atom. Nucl. {\bf 56 }, 1795 (1993) [Yad. Fiz. {\bf 56}, 2137 (1993)];
hep-ph/9911237;
A.Yu. Dubin, A.B.Kaidalov, Yu.A.Simonov,
Phys. Lett. {\bf B323}, 41 (1994).
%
\bibitem{23} H.G. Dosch, Phys. Lett. B {\bf 190}, 177 (1987);
Yu.A. Simonov, Nucl. Phys. B {\bf 307}, 512 (1988);
H.G. Dosch and Yu.A. Simonov, Phys. Lett. B {\bf 205}, 339 (1988).
%
\bibitem{24}A.M. Badalian, B.L.G. Bakker, Yu.A. Simonov,
Phys. Rev. D {\bf 66}, 034026 (2002);
A.M. Badalian, B.L.G. Bakker, Phys. Rev. D {\bf 66} (2002) 034025.
%
\bibitem{25}
A.M. Badalian, V.L. Morgunov, Phys. Rev. D {\bf 60},116008 (1999);
A.M. Badalian, B.L.G. Bakker, Phys. Rev. D {\bf 62}, 094031
(2000).
%
\bibitem{26} Yu.S. Kalashnikova, A.V. Nefediev, Yu.A. Simonov,
Phys. Rev. D {\bf 64}, 014037 (2001);
Yu.S. Kalashnikova and A.V.Nefediev, Phys. Lett. B {\bf 530}, 117 (2002);
A.V. Nefediev, JETP Lett. {\bf 78}, 349 (2003);
Yu.S. Kalashnikova, A.V. Nefediev,
Phys. At. Nucl. {\bf 60}, 1389 (1997); ibid. {\bf 61}, 785 (1998).
%
\bibitem{27} Yu.A. Simonov, Phys. Atom Nucl. {\bf 67}, 1027 (2004)
[Yad. Fiz. {\bf 67}, 1050 (2004);
Yu.A. Simonov (in preparation).
%
\bibitem{28}
Yu.A. Simonov, Phys. Atom. Nucl {\bf 60}, 2069 (1997),
hep-ph/9704301;
Yu.A. Simonov and J.A. Tjon, Phys. Rev. D {\bf 62}, 014501
(2000), ibid {\bf 62},  094511 (2000);
Yu.A. Simonov, Phys. Rev. D {\bf 65}, 094018 (2002).
%
\bibitem{29} V.A. Fock, Izv. Akad. Nauk USSR, OMEN (1937), 557;
J. Schwinger, Phys. Rev. {\bf 82}, 664 (1951);
R.P. Feynman, Phys. Rev. {\bf 80}, 440 (1950); ibid. {\bf 84}, 108 (1951).
%
\bibitem{30} Yu.A. Simonov, J.A. Tjon, Ann. Phys. {\bf 228}, 1 (1993);
ibid {\bf 300}, 54 (2002).
%
\bibitem{31} L. Brink, P. Di Vecchia, P. Howe,
Nucl. Phys. {\bf B 118}, 76 (1977);
A. Polyakov, {\it Gauge fields and strings},
Harwood Ac. Publ., 1987.
%
\bibitem{32} B.S. De Witt, Phys. Rev. {\bf 162}, 1195, 1239 (1967);
J. Honerkamp, Nucl. Phys. B {\bf 48}, 269 (1972);
G.'t Hooft Nucl. Phys. B {\bf 62}, 444 (1973),
Lectures at Karpacz, in: Acta Univ. Wratislaviensis {\bf 368}, 345 (1976);
L.F. Abbot, Nucl. Phys. B {\bf 185}, 189 (1981).
%
\bibitem{33} Yu.A. Simonov, in:
Lecture Notes in Physics, v.479, p.144, Springer, 1996;
Yu.A. Simonov, Phys. At. Nucl. {\bf 58}, 107 (1995)
[Yad. Fiz. {\bf 58},  113 (1995)]; hep-ph/9311247.
%
\bibitem{34} Yu.A. Simonov, in
Proceedings of the XVII Autumn School, Lisboa, Portugal, 29 Sept.- 4 Oct.
Eds. L. Ferreira, P. Nogueira and J. Silva-Marco
(World Scientific, Singapore, 2000) p.60; [arXiv:hep-ph/9911237].
%
\bibitem{35} Yu.A. Simonov, Phys. Lett. B {\bf 515}, 137 (2001);
hep-ph/0105141,
see Err. in A. Di Giacomo, Yu.A. Simonov, Phys. Lett. B {\bf 595}, 368 (2004).
%
\bibitem{36}
R.P. Feynman, A.R. Hibbs, {\it Quantum Mechanics and Path Integrals},
McGraw-Hill, N.Y., 1965.
%
\bibitem{37} V.L. Morgunov, A.V. Nefediev and Yu.A. Simonov,
Phys. Lett. B {\bf 459}, 653 (1999).
%
\bibitem{38} A.M. Badalian, A.I. Veselov, B.L.G. Bakker,
Phys. Atom. Nucl. {\bf 67}, 1367 (2004); hep-ph/0311010;
A.M. Badalian, A.I. Veselov,  B.L.G. Bakker,
J. Phys. G {\bf 31}, 417 (2005).
%
\bibitem{39} A.M. Badalian,  D.S. Kuzmenko,
Phys. Rev.D  {\bf 65}, 016004 (2002).
%
\bibitem{40} G. Bali, Phys. Lett. B {\bf 460}, 170 (1999).
%
\bibitem{41} (CDF Collaboration) A. Abulencia et al.,
Phys. Rev. Lett. {\bf 96}, 082002 (2006); hep-ex/0505076;
(CDF and D0 Collaborations) P. Catastini, hep-ph/0605051.
%
\bibitem{42} J. Pantaleone, S.-H.H. Tye, and Y.J. Ng,
Phys. Rev. D {\bf 33}, 777 (1986).
%
\bibitem{43} A.M. Badalian, B.L.G. Bakker,
Phys.Atom. Nucl. {\bf 69}, 734 (2006);
Phys. Rev. {\bf  D64}, 114010 (2001).
%
\bibitem{44} A. Di Giacomo, H. Panagopoulos,
Phys. Lett. B {\bf 285}, 133 (1992);
G.S. Bali, N. Brambilla, A. Vairo, Phys. Lett. B {\bf 421}, 265 (1998).
%
\bibitem{45} M. D'Elia, A. Di Giacomo, E. Meggiolaro,
Phys. Lett. B {\bf 408}, 315 (1997).
%
\bibitem{46} A.M. Badalian, B.L.G. Bakker,
Phys. Rev. D {\bf 67}, 071901 (2003).
%
\bibitem{47} V.V. Kiselev, Central Eur. J. Phys. {\bf 2}, 523 (2004).
%
\bibitem{48} A. Le Yaouanc, et al.,  Z. Phys. C {\bf 28},309 (1985);
Yu.A. Simonov, Phys. At. Nucl. {\bf 64}, 1876 (2001).
%
\bibitem{49} Yu.S. Kalashnikova, D.S. Kuzmenko,
Phys. At. Nucl. {\bf 67}, 538 (2004), ibid {\bf 66}, 955 (2003).
%
\bibitem{50}
A.B. Kaidalov, Yu.A. Simonov, Phys. Lett. B {\bf 477}, 163 (2000);
ibid. {\bf 636} 101 (2006).
%
\bibitem{51} A.M. Badalian,  B.L.G. Bakker, hep-ph/0604243.
%
\bibitem{52} M.A. Shifman, M.B. Voloshin, J. Nucl. Phys. {\bf 45}, 292 (1987).
%
\bibitem{53} K.G. Chetyrkin, A.G. Grozin, Nucl. Phys. B {\bf 666}, 289 (2003).
%
\bibitem{54} Yu.A. Simonov, Phys. Atom Nucl. {\bf 68}, 709 (2005)
[Yad. Fiz. {\bf 68}, 739 (2005)], hep-ph/0407027.

\end{thebibliography}
\end{document}